\documentclass[lettersize, final, twoside, journal]{IEEEtran}
\usepackage{amssymb, amsmath, amsthm, amsfonts}
\usepackage{booktabs}
\usepackage{bm, color}
\usepackage{setspace}
\usepackage{algorithm}
\usepackage{algorithmic}
\usepackage{longtable, threeparttable}
\usepackage{graphicx}
\usepackage{array}
\usepackage[caption=false, font=footnotesize]{subfig}
\usepackage{textcomp}
\usepackage{float}
\usepackage{stfloats}
\usepackage{url}
\usepackage{verbatim}
\usepackage{graphicx}
\usepackage{cite}
\usepackage{ragged2e}
\usepackage{fancyhdr}
\usepackage{bbding}
\usepackage{makecell}
\usepackage{multirow}
\usepackage{dsfont}
\usepackage{datetime}
\usepackage{footnote}
\usepackage{siunitx}
\usepackage{framed}

\newtheorem{lemma}{Lemma}
\newtheorem{theorem}{Theorem}
\newtheorem{remark}{Remark}
\newtheorem{corollary}{Corollary}
\newtheorem{proposition}{Proposition}
\newtheorem{assumption}{Assumption}

\allowdisplaybreaks[4]

\makeatletter

\newcommand{\Rmnum}[1]{\expandafter\@slowromancap\romannumeral #1@}
\makeatother

\allowdisplaybreaks[4]
\begin{document}

\title{Decentralized Federated Learning with Asynchronous Parameter Sharing for Large-scale IoT Networks}

\author{Haihui Xie, Minghua~Xia,~\IEEEmembership{Senior Member,~IEEE}, Peiran Wu,~\IEEEmembership{Member,~IEEE}, \\ Shuai~Wang,~\IEEEmembership{Member,~IEEE}, and Kaibin Huang,~\IEEEmembership{Fellow,~IEEE}
\thanks{Manuscript received 18 June 2023; revised 2 August 2023 and 24 December 2023; accepted 12 January 2024. This work was supported in part by the National Natural Science Foundation of China under Grants U2001213, 62171486, and 62371444, and in part by the Research Grants Council of the Hong Kong Special Administrative Region, China, under a fellowship award (HKU RFS2122-7S04), the Areas of Excellence scheme grant (AoE/E-601/22-R), Collaborative Research Fund (C1009-22G), and the Grant 17212423. The work was also supported by the Guangdong Basic and Applied Basic Research Foundation under Grants 2019B1515130003, 2021B1515120067, and 2022A1515140166. \textit{(Corresponding author: Minghua Xia.)}
		
Haihui Xie, Minghua Xia, and Peiran Wu are with the School of Electronics and Information Technology, Sun Yat-sen University, Guangzhou, 510006, China (e-mail: xiehh6@mail2.sysu.edu.cn, \{xiamingh, wupr3\}@mail.sysu.edu.cn). 

Shuai Wang is with the Shenzhen Institute of Advanced Technology, Chinese Academy of Sciences, Shenzhen, China (e-mail: s.wang@siat.ac.cn).

Kaibin Huang is with the Department of Electrical and Electronic Engineering, The University of Hong Kong, Hong Kong (e-mail: huangkb@eee.hku.hk).

Color versions of one or more of the figures in this article are available online at https://ieeexplore.ieee.org.
	
Digital Object Identifier 
}
}
 
\markboth{IEEE Internet of Things Journal} {Xie \MakeLowercase{\textit{et al.}}: Decentralized Federated Learning with Asynchronous Parameter Sharing}

\maketitle

\IEEEpubid{\begin{minipage}{\textwidth} \ \\[12pt] \centering 2327-4662 \copyright\ 2022 IEEE. Personal use is permitted, but republication/redistribution requires IEEE permission. \\
See \url{https://www.ieee.org/publications/rights/index.html} for more information.\end{minipage}}

 \IEEEpubidadjcol
 
\begin{abstract}
Federated learning (FL) enables wireless terminals to collaboratively learn a shared parameter model while keeping all the training data on devices per se. Parameter sharing consists of synchronous and asynchronous ways: the former transmits parameters as blocks or frames and waits until all transmissions finish, whereas the latter provides messages about the status of pending and failed parameter transmission requests. Whatever synchronous or asynchronous parameter sharing is applied, the learning model shall adapt to distinct network architectures as an improper learning model will deteriorate learning performance and, even worse, lead to model divergence for the asynchronous transmission in resource-limited large-scale Internet-of-Things (IoT) networks. This paper proposes a decentralized learning model and develops an asynchronous parameter-sharing algorithm for resource-limited distributed IoT networks. This decentralized learning model approaches a convex function as the number of nodes increases, and its learning process converges to a global stationary point with a higher probability than the centralized FL model. Moreover, by jointly accounting for the convergence bound of federated learning and the transmission delay of wireless communications,  we develop a node scheduling and bandwidth allocation algorithm to minimize the transmission delay. Extensive simulation results corroborate the effectiveness of the distributed algorithm in terms of fast learning model convergence and low transmission delay. 
\end{abstract}


\section{Introduction}
\IEEEPARstart{T}{he}  phenomenal development of wireless networking technologies yields a volume of various sensory data, such as traffic information and industrial Internet of Things (IIoT) data. As massive amounts of data are typically required to train a machine learning (ML) model, e.g., convolutional neural networks (CNN), traditional ML algorithms must gather training data from edge devices. Moreover, to realize data sharing, it is essential to transmit a large dataset to an edge server for training ML models. However, limited communication resources for data transmission and conveying large volumes of data by edge devices may substantially burden wireless communications, regardless of privacy constraints. Therefore, edge learning has been proposed to train local models over distributed edge nodes, allowing edge nodes to contribute to the learning procedure for consensus without any raw data exchange. More specifically, federated learning (FL) is, in essence, collaborative edge learning, which deploys large data over multiple edge nodes and then collaborates in training local models without sharing their possibly private data. Moreover, various network architectures regarding computational models and communication mechanisms exist to realize efficient computation and reliable communications.

 \IEEEpubidadjcol
 
\subsection{Related Works and Motivation}
\begin{figure}[t!]
	\centering
	\subfloat[The centralized FL model.]{\includegraphics[width=0.2\textwidth]{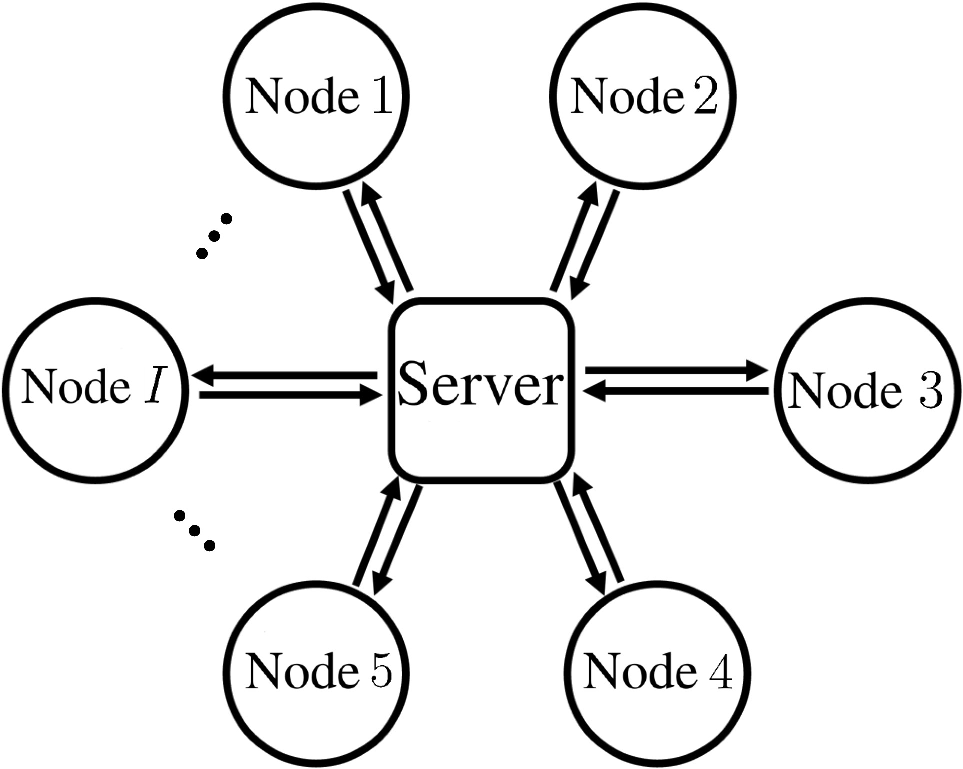}\label{S1-Fig1a}}
	\hfil
	\subfloat[The decentralized FL model.]{\includegraphics[width=0.2\textwidth]{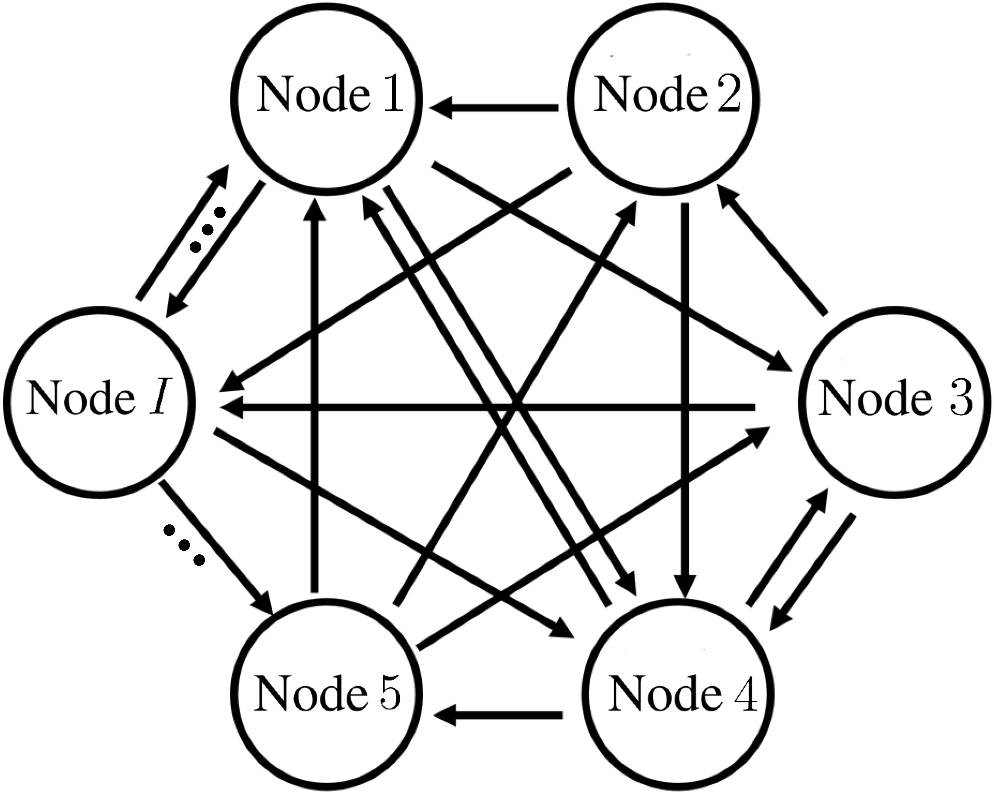}\label{S1-Fig1b}}
	\hfil
	\caption{The conventional centralized and decentralized FL models \cite{Dai0H0T22}.}
	\label{S1-Fig1}
\end{figure}

From the perceptive of computational models, there are two main categories: centralized and decentralized, as shown in Fig.~\ref{S1-Fig1}. Specifically, the federated averaging model is a typical centralized model with a central server to orchestrate a training process, as shown in Fig.~\ref{S1-Fig1a}. After receiving local models from the participating nodes, the central server is responsible for parameter aggregation \cite{ZhangSFYA21}. However, the centralized model requires massive communications between the edge server and nodes, which is usually not feasible. To deal with this infeasibility,  decentralized FL models have recently emerged to significantly reduce the transmission delay because all nodes only communicate with their neighbors, as shown in Fig.~\ref{S1-Fig1b}. For example, the work \cite{Dai0H0T22} optimizes the linking weights of peer-to-peer communications to realize efficient communication in the decentralized scheme. The work \cite{EsfandiariTJBHH21} exploits the cross gradients and develops a novel decentralized learning algorithm, which reduces the communication cost and maintains a satisfactory learning performance for various data distributions.

From the perceptive of communication mechanisms, there are synchronous transmission \cite{8664630, Scheduling2020Yang} and asynchronous transmission \cite{Adaptive2021Lee, ChenSJ20}. For a synchronous scheme, the server cannot aggregate local models until participating nodes finish uploading their local models. Specifically, the works \cite{8664630, Scheduling2020Yang} adopt synchronous FL architectures to balance resource budget and learning accuracy adaptively. By using dynamic aggregation frequency \cite{8664630} or node scheduling \cite{Scheduling2020Yang}, these architectures could address resource-limited problems in the learning process. Nevertheless, while multiple nodes have extensive data for model training or suffer from high transmission delays, synchronous schemes must wait for uploading models. Moreover, the waiting time leads to idling and wastage of computational resources. To address these issues, asynchronous schemes are proposed. The works \cite{Adaptive2021Lee, ChenSJ20} design asynchronous FL systems to reduce transmission delay. However, these asynchronous schemes may yield objective inconsistency, i.e., the learning model may be far from the desired one. The inconsistency between local and global models potentially deteriorates convergence and, even worse, leads to model divergence as the number of iterations increases. In addition, asynchronous schemes often lack theoretical identifications to guarantee convergence, particularly for non-convex FL models.

From the perspective of wireless resource allocation, many works have been on designing communication-efficient FL algorithms that enable edge nodes to efficiently train and transmit model parameters and significantly improve learning performance and convergence rate. For instance, the works \cite{8952884, 9237168} focus on joint scheduling and bandwidth allocation algorithm designs while guaranteeing learning performance. To characterize the theoretical relations between the number of communication rounds and learning performance, the work \cite{8737464} made a tradeoff for joint optimization, including communication latencies and energy consumption. Hence, realizing efficient communication and minimizing transmission delay is essential for distributed algorithms in large-scale IoT networks.

This paper develops a decentralized FL model with asynchronous parameter sharing to address the abovementioned issues. In particular, the proposed model has multi-branch architectures and allows various learning parameters at each node, forming a sum of multi-branch non-convex functions. By recalling the Shapley-Folkman lemma \cite{BiT20}, a sum of multi-branch non-convex functions can approach a convex function when the number of branches becomes large. Thus, the decentralized FL model is less non-convex, and its learning process converges to a global stationary point with a higher probability than the centralized FL model.	In addition, when the dataset is uniformly deployed over each node, and the size of the dataset becomes large, the learning performance and convergence rate of training algorithms improve with the number of distributed nodes increases.\footnote{In principle, the assumption requires that the data distribution at each node is identical to the distribution of the total dataset \cite{180601845}, which is unfortunately hard to satisfy in real-world applications. To address this issue, the cross-gradient aggregation (CGA) algorithm is widely used in simulation experiments to approach the distribution of the total dataset \cite{EsfandiariTJBHH21}.} On the one hand, the asynchronous parameter-sharing mechanism makes the learning and parameter transmission occur in parallel. Thus, a high learning efficiency can be obtained compared with a synchronous architecture. On the other hand, designing a decentralized FL model at the distributed nodes requires an efficient communication network that balances learning performance maximization and transmission delay minimization. In this regard, the asynchronous parameter-sharing method is adopted among distributed nodes. Accordingly, we derive a convergence bound incorporating transmission delay and learning performance. In light of the convergence bound and aggregated duration, we also develop algorithms regarding node scheduling and bandwidth allocation to minimize the transmission delay while maximizing the learning performance. In brevity, Table~\ref{Table0} compares the state-of-the-art and proposed schemes.

\begin{table*}[t!]
	\centering
	\renewcommand\arraystretch{1.25}
	\caption{Comparison of The State-of-the-arts and Proposed Schemes.}
	\vspace{-5pt}
	\label{Table0}
	\begin{threeparttable}[!t]
		\resizebox{\linewidth}{!}{
			\begin{tabular}{!{\vrule width1.2pt} c !{\vrule width1.2pt} c|c|c|c|c|c|c|c!{\vrule width1.2pt}}
				\Xhline{1.2pt} 
				\textbf{Type\tnote{a}} & \textbf{Scheme} &  
				\textbf{\begin{tabular}[c]{@{}c@{}}Learning \\ Accuracy\tnote{b} \end{tabular}} 
				& \textbf{\begin{tabular}[c]{@{}c@{}}Convergence Rate\tnote{c} \end{tabular}}
				& \textbf{\begin{tabular}[c]{@{}c@{}}Less \\ Non-convex\tnote{d} \end{tabular}}
				&  \textbf{\begin{tabular}[c]{@{}c@{}}Cross-gradient \\ Aggregation \end{tabular}}
				&
				\textbf{\begin{tabular}[c]{@{}c@{}} Learning \\ Efficiency\tnote{e} \end{tabular}}
				& \textbf{\begin{tabular}[c]{@{}c@{}}Communication \\ Round\end{tabular}}
				& \textbf{\begin{tabular}[c]{@{}c@{}}Communication \\ Mode\tnote{f} \end{tabular}}
				\\ \Xhline{1.2pt} 
				\multirow{2}{*}{\textbf{\begin{tabular}[c]{@{}c@{}} Cen. \end{tabular}}} 
				& \cite{ZhangSFYA21}          &   ++       & $\mathcal{O} \left( 1 / J \right)$   & \XSolidBrush                            & \XSolidBrush         & $1 / \left( \Gamma + 1 \right)$  & $\mathcal{O} \left( J \right)$                                                                  & Syn. 
				\\ \cline{2-9}
				& \cite{8664630,Adaptive2021Lee}   &  +              & $\mathcal{O} \left( I / J \right)$       & \XSolidBrush         & \XSolidBrush     & $1 / 2$ & $\mathcal{O} \left( J / \Gamma \right)$   &    Syn.                                                   \\ \hline
				\multirow{4}{*}{\textbf{\begin{tabular}[c]{@{}c@{}} Decen. \end{tabular}}}  
				& \cite{Dai0H0T22}   &   ++    &     $\mathcal{O} \left( 1 / J \right)$ & \XSolidBrush                                                                  & \XSolidBrush  & $\Gamma / \left( \Gamma + 1 \right)$                                                                                               & $\mathcal{O} \left( J / \Gamma \right)$  & Asyn.
				\\ \cline{2-9}
				& \cite{EsfandiariTJBHH21}   & +++ & $\mathcal{O} \left( 1 / J + 1 / \sqrt{I J} + 1 / {J}^{2} \right)$ & \XSolidBrush & \Checkmark & $\Gamma / \left( \Gamma + 1 \right)$ & $\mathcal{O} \left( J / \Gamma \right)$ & Asyn.
				\\ \cline{2-9}
				& \textbf{Proposed}                                                & +++                                                                                                                          & $\mathcal{O} \left( 1 / (I J) \right)$    & \Checkmark                         & \Checkmark                         & $\Gamma / \left( \Gamma + 1 \right)$  &  $\mathcal{O} \left( J / \Gamma \right)$  & Asyn.
				\\ \hline
				\Xhline{1.2pt} 
			\end{tabular}
		}
		{\scriptsize
			\begin{tablenotes}
				\item[a] Cen./Decen.: centralized/decentralized FL models;
				\item[b] The symbols ``+, ++, +++'' indicate low, moderate, and high capability, respectively;
				\item[c] $I$: the number of nodes, $J$: the number of iterations, $\Gamma$: the number of aggregated durations;
				\item[d] The tick ``\Checkmark'' indicates a functionality supported, whereas the cross ``\XSolidBrush" indicates not supported;
				\item[e] The learning efficiency is defined as the ratio of the learning time to the total time;
				\item[f] Syn./Asyn.: synchronous/asynchronous communication modes.					
			\end{tablenotes}
		}	
	\end{threeparttable}
\end{table*}

\subsection{Summary of Main Results}
This paper starts to develop a decentralized FL model with non-convex loss functions. Then, we characterize the duality gap of the decentralized FL model by using the seminal Shapley-Folkman lemma. Next, we design a distributed algorithm with asynchronous parameter sharing and derive its convergence bound, where the transmission delay is associated with the convergence bound. At last, we design a node scheduling and bandwidth allocation algorithm to minimize the transmission delay. In brief, the major contributions of the paper are summarized as follows:
\begin{itemize}
	\item[1)] A normalized duality gap is analytically attained to measure the degree of non-convexity for the decentralized FL model. This duality gap is essential to obtain a convex relaxation on loss functions, reducing the influence of non-convexity and making convergence faster.
	\item[2)] A distributed algorithm is developed to learn our decentralized model, including a gradient descent algorithm and asynchronous parameter-sharing principles. Specifically, asynchronous parameter sharing mitigates the influence of stochastic transmission delays.
	\item[3)] The transmission delay is minimized to enhance the convergence rate via efficient node scheduling and bandwidth allocation. By the convergence bound and the transmission delay, minimizing the transmission delay improves learning performance.
\end{itemize}

\subsection{Organization}
The rest of this paper is organized as follows. Section~\ref{Section-SystemModel} formulates the decentralized learning problem. Section~\ref{Section-DualityGap} derives a normalized duality gap between the primal and dual problems. Section~\ref{Section-AlgorithmDesign} develops a distributed algorithm with asynchronous parameter sharing and derives its convergence bound. Section~\ref{Section-ResourceAllocation} establishes the relationship between the transmission delay and convergence bound and allocates radio resources to minimize the transmission delay effectively. Section~\ref{Section-Simulation} presents and discusses the experimental results; finally, Section~\ref{Section-Conclusions} concludes the paper.

{\it Notation}:
Scalars and vectors are denoted by italic regular letters and lower-case ones in bold typeface, respectively. Both space and sets are denoted by calligraphic letters and can be easily identified in the context. The symbol $\bm{0}$ indicates a column vector with all-zero elements. The operation $\left| \mathcal{X} \right|$ denotes the cardinality of the set $\mathcal{X}$, and $\mathcal{Y} \setminus \mathcal{X}$ denotes the complement of set $\mathcal{Y}$ except $\mathcal{X}$. The arithmetic operations $\bm{x} \succeq \bm{y}$ denotes that each element of $\bm{x}$ is greater than or equal to the counterpart of $\bm{y}$. The Landau notation $\mathcal{O} ( \cdot )$ denotes the order of arithmetic operations. The notation $U (x_{0}, \, x_{1})$ denotes a random variable uniformly distributed over the interval $\left[ x_{0}, \, x_{1} \right]$, $x_{0} \leq x_{1}$. Moreover, the ceiling operator is defined as $\left\lceil x \right\rceil \triangleq \min\{ n \in \mathbb{Z} : n \geq x \}$, where $\mathbb{Z}$ is the set of integers. The operator $\left[ \cdot \right]^{+}$ denotes the orthogonal projection of $\bm{w}_{i}$. The functions $\sup ( \cdot )$ and $\inf ( \cdot )$ take a supermum and an infimum, respectively. Finally, we define a binary indicator function $\mathbb{I} (x) = 1$ if the condition $x$ is true and $\mathbb{I} (x) = 0$, otherwise. The main symbols used throughout the paper are defined in Table~\ref{S1-TA1} for ease of reference. 
\begin{table*}[t!]
	\small
	\centering
	\caption{Summary of main notations}
	\begin{tabular}{!{\vrule width1.2pt}c !{\vrule width1.2pt} c !{\vrule width1.2pt}c !{\vrule width1.2pt}c !{\vrule width1.2pt}c !{\vrule width1.2pt}}
		\Xhline{1.2pt}
		\Xhline{1.2pt}
		\textbf{Symbol} & \textbf{Definition} \\
		\Xhline{1.0pt}
		\Xhline{1.0pt}
		$f$; $r$ & The loss function at a training data sample; the regularization function. \\			
		$\bm{w}_{i|{\rm sh}}$ & The sharing parameters of node~$i$ from other nodes. \\
		$\bm{w}_{i}$; $\bm{w}$ & The specific parameters of node~$i$; the current aggregated parameter. \\
		$\bm{v}_{i}$ & The aggregated parameter of node~$i$, which is delayed. \\
		$\bm{s}_{i}$ & The update descent gradient of node~$i$. \\
		$d_{j, \, i}$ & The Euclidean distance between node $i$ and node $j$. \\
		$h_{j, \, i}$ & A small-scale fading with one mean power from node $i$ to node $j$. \\
		$\kappa$; $q_{i}$; $S$ & The learning efficiency; the sparsity ratio; the parameter quantitative bit number. \\
		$B_{i}$; $B$ & The transmission bandwidth from node~$i$ to node~$j$; the total bandwidth. \\
		$F_{i}; \, \hat{F}_{i}$ & Two local loss functions of the centralized and decentralized models, respectively. \\
		$\tilde{F}_{i}$; $\bar{F}$ & The convex relaxation function; the total loss function of all nodes. \\				
		$P$; $T$ & The transmit power; the average traning latency in each iteration. \\
		${\rm SINR}_{j, \, i}$; $R_{i}$ &  The SINR from node~$i$ to node~$j$; an average uplink data rate. \\
		${J}_{i, \, j}$; ${Q}_{i, \, j}$; $Y_{i}$ &  The received state; the scheduled state; the node set. \\
		$\tau_{i, \, j}$ & The communication timestamp from node~$j$ to node~$i$. \\
		$\alpha_{i}$ & A certain fraction of local models that are involved in aggregation. \\
		$\gamma$; $\sigma^{2}$ & The SINR threshold; the variance of Gaussian noise. \\
		$\bm{\xi}_{j}$ & An arbitrary training sample in a dataset. \\
		$\bm{\lambda}$; $\lambda_{i}$ & The total dual variable; the dual variable of node $i$. \\
		$\mathcal{Y}_{i}$; $\Gamma$ & The scheduled nodes set of node~$i$; the aggregated duration. \\
		$\Phi_{j}$ & The out of cell nodes that interferes with node $j$. \\
		$\Gamma_{W}$ & The waitting duration reserved for unscheduled nodes. \\
		$\Gamma_{T}$ & The maximized transmission duration for scheduled nodes. \\
		$\Gamma_{j, \, i}$ & The transmission duration from node $i$ to node $j$. \\
		$\mathcal{D}_{i}$; $\mathcal{I}$ & The local dataset of node~$i$; the index set $\{1, \, 2, \, \cdots, \, I\}$. \\
		$\bar{\mathcal{I}}$ & The nodes set scheduled in this wireless network. \\
		$\mathcal{W}_{i}$; $\mathcal{W}$; $\mathcal{Q}$ & The Euclidean spaces; stacked space; the dual function. \\			
		$\nabla$; $\nabla_{i}$ & The derivative of the aggregated and $i^{\rm th}$ variables, respectively. \\
		$\Delta_{i}$ & The divergence between a function value and its convex relaxation at node $i$. \\
		$\Delta_{\rm worst}$ & The possible maximal divergence of all models. \\
		\Xhline{1.2pt}
		\Xhline{1.0pt}
	\end{tabular}
	\label{S1-TA1}
\end{table*}

\begin{figure}[t!]
	\centering
	\includegraphics[width=0.375\textwidth]{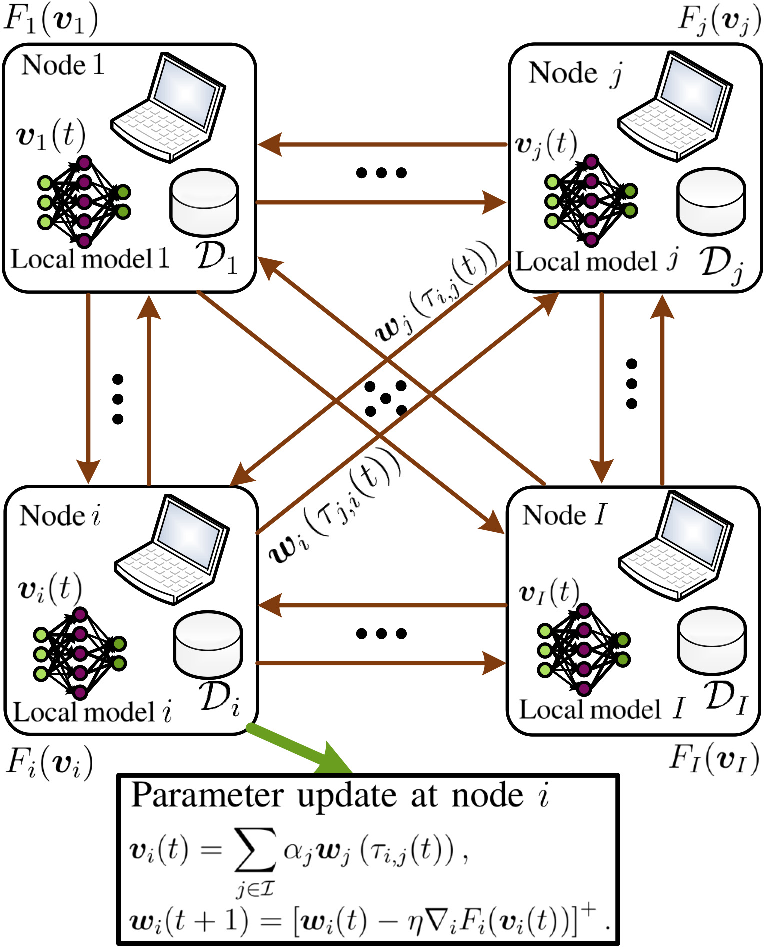}
	\caption{The proposed decentralized FL model with an asynchronous parameter-sharing.}
	\label{S2-Fig1}
	\vspace{-10pt}
\end{figure}

\section{System Model and Problem Formulation} \label{Section-SystemModel}
In this section, we develop a decentralized FL model with an asynchronous parameter-sharing transmission. As shown in Fig.~\ref{S2-Fig1}, the decentralized FL model involves $I$ distributed nodes, local datasets $\left\{\mathcal{D}_{1}, \, \mathcal{D}_{2}, \, \cdots, \, \mathcal{D}_{I}\right\}$, local parameters $\left\{\bm{w}_{1}, \, \bm{w}_{2}, \, \cdots, \, \bm{w}_{I} \right\}$, local aggregated parameters $\left\{\bm{v}_{1}, \, \bm{v}_{2}, \, \cdots, \, \bm{v}_{I} \right\}$, and local loss functions $\left\{F_{1} (\bm{v}_{1}), \, F_{2} (\bm{v}_{2}), \, \cdots, \, F_{I} (\bm{v}_{I})\right\}$, where $\bm{v}_{i}, \,\forall i \in \mathcal{I}$ denotes a locally aggregated parameter at node $i$. For model learning, node $i$ transmits $\bm{w}_{i} \left(\tau_{j, \, i}(t)\right)$ to node $j$ at the $\left(\tau_{j, \, i}(t)\right)^{\rm th}$ iteration, and then node $j$ learns an independent loss function $F_{j} (\bm{v}_{j})$ while receiving $\bm{w}_{i} \left(\tau_{j, \, i}(t)\right)$ at the $t^{\rm th}$ iteration, where $\tau_{j, \, i} (t)$ is a transmission iteration before the $t^{\rm th}$ iteration. Clearly, there exists a transmission period $t - \tau_{j, \, i} (t)$ during parameter transmission between nodes $i$ and $j$. Meanwhile, node $i$ can obtain an aggregated parameter $\bm{v}_{i} (t)$ while receiving all $\bm{w}_{j} (\tau_{i, \, j} (t))$ from other nodes $j \in \mathcal{I} \setminus i$. Moreover, Fig.~\ref{S2-Fig1} also shows that the number of nodes involved in the decentralized learning model is associated with the number of learning parameters $\bm{w}_{i}$, and hence the proposed model with the summation function involves each independent $\bm{w}_{i}$. From the perspective of non-convex geometric analysis, the sum of multiple non-convex functions approaches a convex function \cite{BiT20, 180601845}. Therefore, as reproduced below, the decentralized model becomes less non-convex than the centralized FL model by recalling the Shapley-Folkman lemma.
\begin{lemma}[The Shapley-Folkman lemma\cite{BiT20}]
	Let $S_{1}, \, S_{2}, \, \cdots, \, S_{n}$ be subsets of $\mathcal{R}^{m}$. For each $z \in {\rm conv} \left( \sum_{i = 1}^{n} S_{i} \right) = \sum_{i = 1}^{n} ({\rm conv} S_{i})$, there exist points $z_{i} \in {\rm conv} S_{i}$ such that $z = \sum_{i = 1}^{n} z_{i}$ and $z_{i} \in S_{i}$ except for at most $m$ values of $i$.
\end{lemma}

Motivated by the fact that different learning parameter have significant effects on each local model in multi-task learning \cite{SmithCST17, NEURIPS2018_432aca3a, MaZCLHC19, 10120724}, we extend a centralized FL model to a decentralized model parameterized by $\{ \bm{w}_{i}, \bm{w}_{i|{\rm sh}} | i \in \mathcal{I} \}$, in which $\bm{w}_{i}$ denotes a learning parameter and $\bm{w}_{i|{\rm sh}}$ a sharing parameter at node~$i$. Moreover, $\bm{w}_{i|{\rm sh}}$ is updated from other nodes and then restricts the amount of local deviation relative to the desired model, defined as
\begin{equation}
\bm{w}_{i|{\rm sh}} (t) \triangleq \dfrac{1}{1 - \alpha_{i}} \sum_{ j \in \mathcal{I} \setminus i } \alpha_{j} \bm{w}_{j} \left( \tau_{i, \, j} ( t ) \right).  \label{S4-EQ4a}
\end{equation}	

In this regard, it is essential to formulate a decentralized FL problem, and we define a loss function of local dataset $\mathcal{D}_{i}$ as
\begin{equation}
\begin{aligned}
& F_{i} \left( ( 1 - \alpha_{i} ) \bm{w}_{i|{\rm sh}} + \alpha_{i} \bm{w}_{i} \right) \\
& \triangleq \dfrac{1}{ | \mathcal{D}_{i} | } \sum_{ j \in \mathcal{D}_{i} } f ( ( 1 - \alpha_{i} ) \bm{w}_{i|{\rm sh}} + \alpha_{i} \bm{w}_{i} \vert \bm{\xi}_{j} ), 
\end{aligned}
\end{equation}
where $\alpha_{i} \triangleq | \mathcal{D}_{i} | / | \mathcal{D} |$ denotes a certain fraction of local models that are involved in aggregation; $\bm{w}_{i|{\rm sh}}$ is independent of the local dataset $\mathcal{D}_{i}$ but changes as the number of the iteration increases\footnote{The shared parameter $\bm{w}_{i | {\rm sh}}$ is initialized by the aggregated parameter of nodes $\mathcal{I} \setminus {i}$, thus it depends on the local datasets $\mathcal{D}_{j}, \, j \in \mathcal{I} \setminus {i}$ but is independent of the local dataset $\mathcal{D}_{i}$ at the learning process.}; $\bm{\xi}_{j}$ is an arbitrary training sample in $\mathcal{D}_{i}$; and $f ( \cdot \vert \bm{\xi}_{j} )$ denotes a non-convex loss function given the training data sample $\bm{\xi}_{j}$. In practice, $f ( \cdot \vert \bm{\xi}_{j} )$ can be a linear regression function, a logistic regression function, a support vector machine function, or a cross-entropy function, among others \cite{9484767}. Now, we can formulate a globally decentralized FL problem as
\begin{subequations}
	\begin{align}
	\mathcal{P}1 : &\min_{\bm{\omega} \in \mathcal{W}} \, \sum\limits_{ i = 1 }^{I} \alpha_{i} F_{i} \left( ( 1 - \alpha_{i} ) \bm{w}_{i|{\rm sh}} + \alpha_{i} \bm{w}_{i} \right) \label{EQ-S2-2a} \\
	&{\rm s.t.} \quad r ( \bm{w}_{i} ) \leq K_{i}, \, \forall \bm{w}_{i} \in \mathcal{W}_{i}, \, i \in \mathcal{I}, \label{EQ-S2-3b}
	\end{align}
\end{subequations}
where $\bm{\omega} \triangleq \left[ \bm{w}_{1}^{T}, \,  \bm{w}_{2}^{T}, \, \cdots, \, \bm{w}_{I}^{T} \right]^{T}$ is a stacked vector, $\mathcal{W} = \mathcal{W}_{1} \times \mathcal{W}_{2} \times \cdots \times \mathcal{W}_{I}$ is a direct-product space involved in $I$ Euclidean sub-spaces $\{\mathcal{W}_{i}\}_{i \in \mathcal{I}}$, $K_{i}$ is an arbitrary value which makes the solution of $\mathcal{P}1$ feasible, and the regularization function $r ( {\bm{w}_{i}} ) \geq 0$ can be chosen from the sparsity constraint $\|\bm{w}_{i}\|_{1}$ or the weight decay $1/ 2 \|\bm{w}_{i}\|_{2}^{2}$ \cite{BeckT09}. Moreover, \eqref{EQ-S2-3b} denotes a soft constraint in which the corresponding hard constraint is replaced by regularization. This replacement can control the complexity of a neural network model to avoid over-fitting and restrict the divergence of parameters arising due to heterogeneity in the data distribution \cite{9252927}.
\begin{remark}
	The proposed decentralized model can attain a higher learning efficiency and less non-convex FL model than the centralized one since the former makes the learning and data exchange parallel. Also, the latter with cell-interior scheduling leads to data deficiency since the resultant model training fails to exploit data at cell-edge devices\cite{8870236}. To address this issue, the proposed method uses the shared parameter with the desired transmission delay and then improves the data quantity.
\end{remark}

So far, we have formulated a decentralized FL problem. By recalling the seminal Shapley-Folkman lemma, we can infer that the sum result in the objective function of $\mathcal{P}1$ becomes more and more convex as the number of nodes (i.e., $I$) increases. Therefore, designing algorithms to solve $\mathcal{P}1$ efficiently improves the FL convergence rate. 

\section{Duality Gap of The Decentralized FL Model} \label{Section-DualityGap}
Although we have formulated a multi-branch problem, building a direct relationship between the node distribution and loss function takes time and effort. To address this issue, this section derives a normalized duality gap between the decentralized FL problem and its dual one to measure the degree of non-convexity. To start with, the dual problem of $\mathcal{P}1$ is given by
\begin{subequations}
	\begin{align}
	\mathcal{P}2 : \max_{ \bm{\lambda} \succcurlyeq \bm{0} } \, & \mathcal{Q} ( \bm{\lambda} ) - \sum_{ i = 1 }^{I} \lambda_{i} K_{i} \\
	{\rm s.t.} \quad & \mathcal{Q} ( \bm{\lambda} ) = \inf_{ \bm{\omega} \in \mathcal{W} } \sum\limits_{ i = 1 }^{I} \left( \alpha_{i} F_{i} \left( ( 1 - \alpha_{i} ) \bm{w}_{i|{\rm sh}} + \alpha_{i} \bm{w}_{i} \right) \right. \nonumber \\
	& \quad \quad \quad \quad +{} \, \left. \lambda_{i} r ( \bm{w}_{i} ) \right),
	\end{align}
\end{subequations}
where $\bm{\lambda} \triangleq \left[ \lambda_{1}, \, \lambda_{2}, \, \cdots, \, \lambda_{I} \right]^{T}$ is a dual variable. For further proceeding, a convex relaxation of the loss function $F_{i} ( ( 1 - \alpha_{i} ) \bm{w}_{i|{\rm sh}} + \alpha_{i} \bm{w}_{i} )$ is given by \cite{180601845}
\begin{align}
\tilde{F}_{i} ( \tilde{\bm{w}} ) \triangleq & \inf_{\theta_{j}, \, \bm{x}_{j} \in \mathcal{W}_{i}} \left\{ \sum_{j = 1}^{N_{i}} \theta_{j} F_{i} \left( ( 1 - \alpha_{i} ) \bm{w}_{i|{\rm sh}} + \alpha_{i} \bm{x}_{j} \right) \right. \nonumber\\
& \left. \left| \tilde{\bm{w}} = \sum_{j = 1 }^{N_{i}} \theta_{j} \bm{x}_{j}, \, \sum_{ j = 1 }^{N_{i}} \theta_{j} = 1, \, \theta_{j} \geq 0 \right. \right\}, \label{EQ-S3-1}
\end{align}
where $\tilde{\bm{w}}$ denotes an optimal variable of $\tilde{F}_{i} ( \tilde{\bm{w}} )$, $N_{i} \geq 2$ is an arbitrary integer which makes $\tilde{F}_{i} ( \tilde{\bm{w}} )$ significant, and $\bm{x}_{j}$ denotes a sampling variable of the Euclidean space. Also, we use $\tilde{\bm{w}}$ again to rewrite the $i^{\rm th}$ sub-function in $\mathcal{P}1$ as
\begin{equation} \label{EQ-S3-2}
\small \hat{F}_{i} ( \tilde{\bm{w}} ) = \inf_{ \bm{w}_{i} \in \mathcal{W}_{i} } \left\{ F_{i} \left( ( 1 - \alpha_{i} ) \bm{w}_{i|{\rm sh}} + \alpha_{i} \bm{w}_{i} \right) \left| r ( \bm{w}_{i} ) \leq r ( \tilde{\bm{w}} ) \right. \right\}.
\end{equation}
Here, we choose an appropriate $K_{i}$ such that $r (\tilde{\bm{w}}) \geq K_{i}, \, i \in \mathcal{I}$. Combining \eqref{EQ-S3-1} with \eqref{EQ-S3-2}, our conclusion is formalized in the following theorem.
\begin{theorem} \label{S3-T1}
	Let the maximal divergence of all models be $\Delta_{\rm worst} \triangleq \max_{i \in \mathcal{I}} \Delta_{i}$, where $\Delta_{i} \triangleq \sup_{ \tilde{\bm{w}} \in \mathcal{W}_{i} } \left\{ \hat{F}_{i} ( \tilde{\bm{w}} ) - \tilde{F}_{i} ( \tilde{\bm{w}} ) \right\} \geq 0 $ denotes the divergence between the function value of $\hat{F}_{i}$ and its convex relaxation $\tilde{F}_{i}$, the normalized duality gap can be bounded by 
	\begin{equation} \label{S3-T1-EQ3}
	0 \leq \dfrac{ \inf ( \mathcal{P}1 ) - \sup ( \mathcal{P}2 ) }{\Delta_{\rm worst}} \leq 2 \alpha_{\max},
	\end{equation}
	where $\alpha_{\max} \triangleq \max_{i = 1, 2, \cdots, I} \alpha_{i}$.
\end{theorem}
\begin{proof}
	See Appendix~\ref{SA-A}.
\end{proof}

In Theorem~\ref{S3-T1}, although the divergence $\Delta_{i}$ changes from node to node, the maximal divergence $\Delta_{\rm worst}$ among all nodes remains almost the same in case the number $I$ of nodes is large enough. Accordingly, the normalized duality gap $\left( \inf ( \mathcal{P}1 ) - \sup ( \mathcal{P}2 ) \right)  / \Delta_{\rm worst}$ in \eqref{S3-T1-EQ3}  measures the degree of non-convexity for the objective function. When differences in the data distribution among all distributed nodes turn larger, i.e., $\alpha_{\max} \rightarrow 1$, the upper bound of \eqref{S3-T1-EQ3} becomes looser, and the proposed model degenerates into a centralized FL model. In contrast, when a large dataset has a uniform distribution over each node, i.e., $\alpha_{1} = \cdots = \alpha_{I} = \alpha_{\max} \rightarrow 1 / I$, the upper bound of \eqref{S3-T1-EQ3} takes the minimum value, i.e., $2/I$. Clearly, the normalized duality gap decreases linearly with~$I$. In other words, $\mathcal{P}1$ becomes more and more convex as $I$ increases. As a result, the solution of $\mathcal{P}1$ approaches that of $\mathcal{P}2$ as $I \rightarrow + \infty$.

\section{Algorithm Design and Convergence Analysis} \label{Section-AlgorithmDesign}
{By using the duality gap, we can now control the non-convexity of our decentralized FL model by adjusting the number of nodes. The learning process can also converge to a global solution with a high probability.} Next, we develop a distributed algorithm with asynchronous parameter sharing, followed by an analytical upper bound to guarantee convergence.

\subsection{Algorithm Design}
Based on the decentralized FL model shown in Fig.~\ref{S2-Fig1}, we now design a distributed algorithm with asynchronous parameter sharing for learning relative parameters. The following assumptions are prerequisites for convergence.
\begin{assumption}[pp. 483-484 of \cite{Parallel1997Dimitri}] \label{S4-A1}
	Assume that the number of transmission delays $\Gamma$, i.e., how usually all scheduled nodes participate in an asynchronous aggregation, is a positive integer, then the transmission iteration $\tau_{i, \, j} ( t )$ satisfies 
	\begin{equation}  \label{S4-Eq-A1}
	\max \{ t - \Gamma, \, 0 \}  < \tau_{i, \, j} ( t ) \leq t, \, \forall i \neq j, \text{ and } \tau_{i, \, i} (t) = t.
	\end{equation}
\end{assumption}

In light of Assumption~\ref{S4-A1}, it is evident that each variable $\bm{w}_{i}$ and sharing parameter $\bm{w}_{i|{\rm sh}}$ at node $i$ can be updated timely if $\Gamma$ is bounded. However, $\bm{w}_{i|{\rm sh}}$ cannot be updated in parallel as it requires information from other nodes. Therefore, we replace the gradient descent algorithm with asynchronous parameter sharing, yielding	
\begin{align}
\bm{v}_{i} ( t ) &{} = \alpha_{i} \bm{w}_{i} (t) + (1 - \alpha_{i}) \bm{w}_{i|{\rm sh}} (t) \nonumber \\
&{} = \sum_{ j \in \mathcal{I} } \alpha_{j} \bm{w}_{j} \left( \tau_{i, \, j} ( t ) \right), \label{S4-EQ4b}
\end{align}
where $\bm{v}_{i} (t)$, serving as an aggregated parameter, is a linear function of variables $\bm{w}_{j} \left( \tau_{i, \, j} ( t ) \right)$, $\forall j \in \mathcal{I}$. Also, $\bm{w}_{i|{\rm sh}}$ in \eqref{S4-EQ4b} reflects information interactions among nodes and ensures that the preliminary model can be initialized properly by drawing knowledge from multiple local data distributions.

Now, we apply gradient descent algorithms to learn our FL model. At first, $\bm{w}_{i} ( t )$ can be updated by
\begin{equation} \label{S4-EQ5}
\bm{w}_{i} ( t + 1 ) = \left[ \bm{w}_{i} ( t ) - \eta \nabla_{i} F_{i} ( \bm{v}_{i} ( t ) ) \right]^{+},
\end{equation}
where $\nabla_{i} F_{i}$ denotes the derivative of $F_{i}$ with respect to the $i^{\rm th}$ parameter $\bm{w}_{i} (t)$, i.e., $\nabla_{\bm{w}_{i}} F_{i}$. Clearly, \eqref{S4-EQ5} is a gradient projection algorithm\cite[p. 213]{Parallel1997Dimitri}, which generalizes the conventional gradient decent method to the case with complex constraints. Incidentally, the mapping $\left[ \cdot \right]^{+}$, i.e., orthogonal projection of $\bm{w}_{i}$, can be defined as
\begin{equation}
\left[ \bm{w}_{i} \right]^{+} \triangleq \underset{\bm{y}}{{\rm argmin}} \, \| \bm{y} - \bm{w}_{i} \|_{2}, \, \forall \bm{y} \in \mathcal{X}_{i},
\end{equation}	
where $\mathcal{X}_{i} \triangleq  \left\{ \bm{y} \left| r(\bm{y}) \leq K_{i}, \right. \forall \bm{y} \in \mathcal{W}_{i} \right\}$. To update the mapping $[\cdot]^{+}$, the iterative shrinkage-thresholding algorithm with $\ell_{1}$-norm constraint, the weight decay with $\ell_{2}$-norm constraint or other regularized solvers can be exploited \cite{BeckT09, 8416722}. Accordingly, a descent gradient of the distributed algorithm can be computed and given by
\begin{align}
\bm{s}_{i} ( t ) &\triangleq \dfrac{1}{\eta} \left( \bm{w}_{i} ( t + 1 ) - \bm{w}_{i} ( t ) \right) \nonumber \\
&= \dfrac{1}{\eta} \left( \left[ \bm{w}_{i} ( t ) - \eta \nabla_{i} F_{i} ( \bm{v}_{i} ( t ) ) \right]^{+} - \bm{w}_{i} ( t ) \right),
\end{align}
Thus, the local parameter given by \eqref{S4-EQ5} can be rewritten as
\begin{equation} \label{S4-EQ6}
\bm{w}_{i} (t + 1) = \bm{w}_{i} (t) + \eta \bm{s}_{i} (t).
\end{equation}

\begin{figure*}[t!]
	\centering
	\includegraphics[width=0.80\textwidth]{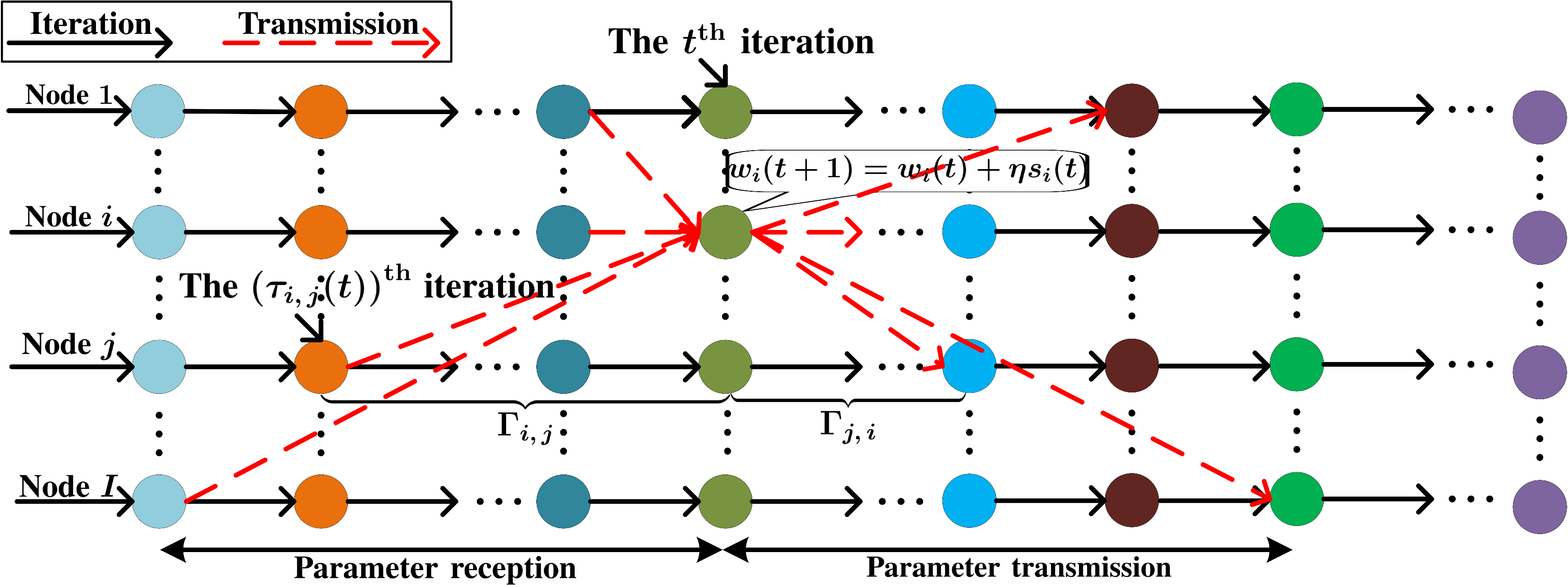}
	\caption{A schematic diagram for iterative learning processes.}
	\label{S4-Fig2}
	\vspace{-10pt}
\end{figure*}

For illustrative purposes, Fig.~\ref{S4-Fig2} shows the schematic diagram for iterative learning processes, where different colored disks denote different learning iterations. By unfolding the time-series data, Fig.~\ref{S4-Fig2} also illustrates the network topology with transmission delays. In particular, parameter reception and transmission are two main processes. In the period of the parameter reception, node $i$ starts receiving the parameters from the other nodes at the $t^{\rm th}$ iteration. Specifically, node $j$ transmits $\bm{w}_{j} \left( \tau_{i, \, j} (t) \right)$ to node $i$ at the $\left( \tau_{i, \, j} (t) \right)^{\rm th}$ iteration, where we take $\Gamma_{i, \, j} \triangleq t - \tau_{i, \, j} (t)$ as a communication duration and the parameter $\bm{w}_{j} \left( \tau_{i, \, j} (t) \right)$ rather than the up-to-date $\bm{w}_{j}(t)$ is used at the $t^{\rm th}$ iteration. After receiving all parameters of the set $\mathcal{I} \setminus i$, node $i$ performs the iteration \eqref{S4-EQ6}. On the other hand, in the period of the parameter transmission, node $i$ transmits the corresponding parameter $\bm{w}_{i} (t + 1)$ to nodes in the set $\mathcal{I} \setminus i$, where we take $\Gamma_{j, \, i}$ as a communication duration between node $i$ and node $j$.

\subsection{Convergence Analysis}
Now, we characterize the convergence for the distributed algorithm with asynchronous parameter sharing. {Compared with traditional convergence analysis, it is hard to prove the convergence of our proposed algorithm due to shared and local learning parameters. To address this issue, this paper designs a new loss function that mitigates the influence of shared parameters and establishes the convergence relationship.} First, we define this loss function as
\begin{equation} \label{S4-EQ8}
\bar{F} \left( \bm{w} \right) \triangleq \sum_{i = 1}^{I} \alpha_{i} F_{i} \left( \bm{w} \right) =\sum_{i = 1}^{I} \alpha_{i} F_{i} \left( \sum_{i = 1}^{I} \alpha_{i} \bm{w}_{i} \right) ,
\end{equation}
where $\bm{w} (t) \triangleq \sum_{i = 1}^{I} \alpha_{i} \bm{w}_{i} (t)$ stands for the last aggregated parameter, unlike $\bm{v}_{i} (t)$ shown in \eqref{S4-EQ4b}. Since the objective function in \eqref{EQ-S2-2a} is indeterminate due to the stochastic transmission delays, \eqref{S4-EQ8} rather than \eqref{EQ-S2-2a} is applied in the following convergence analysis. Then, it is sufficient to require that each variable is updated in a gradient descent direction. Accordingly, the following assumptions are introduced for the convergence analysis. 

\begin{assumption}[pp. 527-528 of \cite{Parallel1997Dimitri}] \label{S4-A2}
	The objective loss and descent gradient satisfy the following conditions:
	\begin{itemize}
		\item[(a)] There holds $F_{i} ( \bm{w} ) \geq 0, \, \forall i \in \mathcal{I}$.
		\item[(b)] Any function $\nabla F_{i}$ is continuously differentiable and there exists a constant $L_{1}$ such that
		\begin{equation} \nonumber
		\| \nabla F_{i} ( \bm{x} ) - \nabla F_{i} ( \bm{y} ) \|_{2} \leq L_{1} \| \bm{x} - \bm{y} \|_{2},
		\end{equation}
		where $\nabla F_{i}(\bm{x})$ (resp., $\nabla F_{i}(\bm{y}$)) represents the derivative with respect to the aggregated variable $\bm{x}$ (resp., $\bm{y}$).
		\item[(c)] For all $i \in \mathcal{I}$, we have 
		\begin{equation} \nonumber
		\bm{s}_{i}^{T} ( t ) \nabla_{i} F_{i} ( \bm{v}_{i} ( t ) ) \leq 0.
		\end{equation}
		\item[(d)] There are positive constants $L_{2}$ and $L_{3}$, such that 
		\begin{equation} \nonumber
		L_{2} \left\| \nabla_{i} F_{i} ( \bm{v}_{i} ( t ) ) \right\|_{2} \leq \left\| \bm{s}_{i} ( t ) \right\|_{2} \leq L_{3} \left\| \nabla_{i} F_{i} ( \bm{v}_{i} ( t ) ) \right\|_{2}.
		\end{equation}
		\item[(e)] There is a positive constant $\delta$, such that 
		\begin{equation} \nonumber
		\| \bm{s}_{i} ( t ) - \bm{s}_{j} ( t ) \|_{2} \leq \delta \min \left( \| \bm{s}_{i} ( t ) \|_{2}, \, \| \bm{s}_{j} ( t ) \|_{2} \right),
		\end{equation}
		where $\delta$ is a convergence bound of $\| \bm{s}_{i} ( t ) - \bm{s}_{j} ( t ) \|_{2}$.
	\end{itemize}
\end{assumption}
Assumption~\ref{S4-A2}(a) means that the loss functions take non-negative values. The smooth assumption given by Assumption~\ref{S4-A2}(b) indicates how fast the gradient of $F_{i} (\bm{x})$ can change. Assumption~\ref{S4-A2}(c) implies that $\bm{s}_{i} (t)$ must be in the same Cartesian quadrant as $-\nabla F_{i} (\bm{v}_{i} (t))$. To avoid gradient exploding and ensure the convergence of model training, Assumption~\ref{S4-A2}(d) is necessary to restrict the scale of gradients. Moreover, Assumption~\ref{S4-A2}(e) means a bounded local dissimilarity and a divergence of the gradients. It is evident that Assumptions~\ref{S4-A2}(d) and \ref{S4-A2}(e) reveal the data heterogeneity and distribution over edge nodes \cite{8664630}. All the assumptions mentioned above characterize the key features of loss and risk functions of existing machine learning models, which have been widely used in the literature about the convergence analysis of machine learning models, even for non-convex models \cite{9252927, 9563947, 9252924}. 

By virtue of Assumption~\ref{S4-A2}, we obtain the following lemma on $\bar{F} (\bm{w})$:
\begin{lemma} \label{S4-L1}
	For any $\bm{w}$ and $\bm{w}_{i}$, we have 
	\begin{equation} \label{S4-EQ13}
	\nabla \bar{F} \left( \bm{w} \right) = \sum_{i = 1}^{I} \nabla_{i} F_{i} \left( \sum_{j = 1}^{I} \alpha_{j} \bm{w}_{j} \right),
	\end{equation}
	and
	\begin{equation} \label{S4-EQ19}
	\| \nabla \bar{F} ( \bm{x} ) - \nabla \bar{F} ( \bm{y} ) \|_{2} \leq L_{1} \| \bm{x} - \bm{y} \|_{2}, \, \forall \bm{x}, \, \bm{y} \in \mathcal{W}.
	\end{equation}
\end{lemma}
\begin{proof}
	By definition, we have
	\begin{equation} \label{S4-EQ14}
	\nabla_{i} F_{i} \left( \sum_{j = 1}^{I} \alpha_{j} \bm{w}_{j} \right) = \alpha_{i} \nabla F_{i} ( \bm{w} ).
	\end{equation}
	Summing up \eqref{S4-EQ14} for $i \in \mathcal{I}$ and inserting it into \eqref{S4-EQ8} yields \eqref{S4-EQ13}. On the other hand, by Assumption~\ref{S4-A2}(b), we get
	\begin{subequations}
		\begin{align}
		&\| \nabla \bar{F} ( \bm{x} ) - \nabla \bar{F} ( \bm{y} ) \|_{2}^{2} \nonumber\\
		&= \left\| \sum_{ i = 1 }^{I} \alpha_{i} \nabla F_{i} ( \bm{x} ) - \sum_{ i = 1 }^{I} \alpha_{i} \nabla F_{i} ( \bm{y} ) \right\|_{2}^{2} \label{S4-L2-EQ19a} \\
		&\leq \sum_{ i = 1 }^{I} \alpha_{i} \left\| \nabla F_{i} ( \bm{x} ) - \nabla F_{i} ( \bm{y} ) \right\|_{2}^{2} \nonumber\\
		&\leq L_{1}^{2} \left\| \bm{x} - \bm{y} \right\|_{2}^{2}, \label{S4-L2-EQ19b} 
		\end{align}
	\end{subequations}
	where \eqref{S4-L2-EQ19a} is due to \eqref{S4-EQ13} and the last inequality of \eqref{S4-L2-EQ19b} is obtained by using $\sum_{ i = 1 }^{I} \alpha_{i} = 1$.
\end{proof}		

Now, we can formalize a convergence bound on how rapid the decrease of global loss is.
\begin{theorem} \label{S5-T2}
	With Assumption~\ref{S4-A2}, there exists a learning rate $\eta > 0$ such that $\bar{F} \left( \bm{w} ( t + 1 ) \right)$ converges to a  bound given by
	\begin{equation} \label{S5-EQ-8}
	\bar{F} \left( \bm{w} ( t + 1 ) \right) \leq \underbrace{ \bar{F} \left( \bm{w} ( 0 ) \right) - \eta u (\eta) \sum_{\tau = 0}^{t} \left\| \bm{g} (\tau) \right\|_{2}^{2} }_{ U (\Gamma) },
	\end{equation}
	where $\bm{g} (\tau) \triangleq \left[ \bm{s}_{1}^{T} (\tau), \, \bm{s}_{2}^{T} (\tau), \, \cdots, \, \bm{s}_{I}^{T} (\tau) \right]^{T}, \, \forall \tau = 0, \, \cdots, \, t$,  represents a stacked global gradient of all nodes, and $u (\eta)$ is defined as
	\begin{equation} \nonumber
	u (\eta) \triangleq \dfrac{I}{L_{3}} - \dfrac{\delta ( I - 1 )}{L_{2}} - \dfrac{\left(I (3 + \eta) - 1 \right) L_{1}}{2} - \dfrac{ ( 3 I - 1 ) \eta^{2} L_{1} \Gamma^{2} }{2}.
	\end{equation}
\end{theorem}
\begin{proof}
	See Appendix~\ref{SA-B}.
\end{proof}	

Theorem~\ref{S5-T2} shows that the learning rate $\eta$ and the number of transmission delay $\Gamma$ are vital factors to determine the upper bound of divergence in our distributed algorithm with asynchronous parameter sharing. Moreover, we can make some exciting corollaries from Theorem~\ref{S5-T2}.
\begin{corollary} \label{S5-C1}
	Theorem~\ref{S5-T2} infers that
	\begin{equation} \label{S5-C1-a}
	\lim\limits_{t \rightarrow + \infty} \bm{w}_{i} (t) = \lim\limits_{t \rightarrow + \infty} \bm{w}_{i} (\tau_{j, \, i} (t)), \, \forall i, \, j \in \mathcal{I},
	\end{equation}
	and 
	\begin{equation} \label{S5-C1-b}
	\lim\limits_{t \rightarrow + \infty} \bm{w} ( t ) = \lim\limits_{t \rightarrow + \infty} \bm{v}_{i} ( t ), \ \forall i \in \mathcal{I}.
	\end{equation}
\end{corollary}
\begin{proof}
	See Appendix~\ref{SA-C}.
\end{proof}	
In essence, \eqref{S5-C1-a} implies that each model can converge independently; \eqref{S5-C1-b} shows that we can obtain an aggregated model at each node, which converges to the same global model. The reason behind these observations is that the sharing parameter $\bm{w}_{i|{\rm sh}}$ can restrict the severity of local deviations if they increase with the number of iterations. 

\begin{corollary} \label{S5-C2}
	Given Assumption~\ref{S4-A2}, there exists a learning rate $\eta > 0$, such that $\bar{F} \left( \bm{w} ( t + 1 ) \right)$ converges to a fixed value. If $ \eta^2 < \delta \ll \eta < 1$, then we obtain
	\begin{equation} \label{Eq-Corollary2}
	\delta < \eta \lessapprox \min \left( \mathcal{O} \left( \dfrac{2}{L_{1} L_{3}} - \dfrac{3 I - 1}{I} \right), \, \sqrt{\delta} \right) .
	\end{equation}
\end{corollary}
\begin{proof}
	Assume that $ \eta^2 < \delta \ll \eta < 1$, then the term $u(\eta)$ defined in \eqref{S5-EQ-8} can be approximated as $u(\eta)\thickapprox {I} / {L_{3}} - {I \eta L_{1}} / {2} - { ( 3 I - 1 ) L_{1} } / {2}$. Since the constraint $u(\eta) > 0$ is satisfied naturally, we attain
	\begin{equation} \label{EQ-AC-C13}
	0 < \eta \lessapprox \mathcal{O} \left( \dfrac{2}{L_{1} L_{3}} - \dfrac{3 I - 1}{I} \right) .
	\end{equation}
	In case of $ \eta^2 < \delta \ll \eta < 1$ and \eqref{EQ-AC-C13}, we can readily obtain \eqref{Eq-Corollary2}.	
\end{proof}	
Corollary~\ref{S5-C2} states that the learning rate $\eta$ is naturally in a reasonable range to guarantee convergence. The convergence rate is prolonged if $\eta$ is small. On the other hand, if $\eta$ is large, the term involving $\eta^{2}$ in the parenthesis of \eqref{S5-EQ-8} cannot be ignored anymore, yielding stochastic fluctuations due to different transmission delays among nodes. In the next section, we aim to minimize the maximum of transmission delays.

\section{Wireless Resource Allocation} \label{Section-ResourceAllocation}
In this section, we develop efficient communications for parameter sharing in highly distributed networks. In particular, it is observed from Theorem~\ref{S5-T2} that minimizing the upper bound is essential to obtain a good convergence in each iteration period. Therefore, it is natural to minimize the convergence bound $U (\Gamma)$ defined in \eqref{S5-EQ-8} concerning the aggregated duration $\Gamma$, where $\Gamma \triangleq \Gamma_{W} + \Gamma_{T}$ includes the number of waiting duration $\Gamma_{W}$ and the maximal transmission duration $\Gamma_{T}$. Specifically, $\Gamma_{W}$ is from the reserved duration for unscheduled nodes \cite{Joint2020Shi}, and $\Gamma_{T}$ is associated with the transmission delay \cite{9170917}. However, as $\Gamma_{W}$ varies due to the randomness of channels, it is hard to express $\Gamma_{W}$ explicitly and provide a corresponding optimization. Therefore, we design a scheduling strategy and a bandwidth allocation algorithm for efficient wireless resource allocation. 	

Like \cite{Scheduling2020Yang, 9435350}, we consider a typical wireless network where all user nodes are uniformly distributed. The network coverage area is tessellated into Voronoi cells using the nearest-neighbor association criteria, i.e., each node is associated with its nearest base station (BS) \cite{8358978}. We assume that the locations of user nodes follow a homogeneous Poisson point process (PPP) with a pre-defined spatial density. A fixed amount of spectrum resource is divided into multiple spectrum bands for efficient spectral utilization. Each node with a single antenna broadcasts the same parameters to others through its allocated spectrum band. A block-fading propagation model is adopted, where the channels between any two nodes are independently and identically distributed (i.i.d.) and keep quasi-static in the parameter transmission period. In addition, the propagation channels are supposed to be narrow-band (which is true in IoT applications), and large-scale path loss and small-scale fading are accounted for as usual.

\subsection{Scheduling Strategy}
In this subsection, we start to analyze the waiting duration $\Gamma_{W}$ during transmission. Using Slivnyak's theorem, it is sufficient to evaluate the communication quality through the signal-to-interference-plus-noise ratio (SINR) at a typical node \cite{Adaptive2021Lee}. For the parameter transmission from node $i$ to $j$, the SINR at node $j$ can be expressed as \cite{8976426}
\begin{equation} \label{S5-EQ22}
{\rm SINR}_{j, \, i} = \dfrac{P h_{j, \, i} d_{j, \, i}^{-\alpha}}{\sum_{x \in \Phi_{j}} P h_{j, \, x} d_{j, \, x}^{-\alpha} + \sigma^{2}},
\end{equation}
where all nodes are assumed to have the same transmit power $P$, the symbol $h_{j, \, i} \sim \exp (1)$ reflects a small-scale fading with unity power (i.e., $\mathbb{E}
\{|h_{j, \, i}|^2\} = 1$), $d_{j, \, i}$ denotes a Euclidean distance between nodes $i$ and $j$, $\alpha$ is a path loss exponent, the set $\Phi_{j}$ stands for out-of-cell nodes interfering with node $j$, and $\sigma^{2}$ denotes the variance of additive white Gaussian noises. 

Due to the limited wireless resources of low-cost IoT nodes, they are divided into distinct groups, each associated with a particular learning task. Accordingly, each node exchanges its learning parameters within the group per se. To obtain good communication qualities in the transmission period, the SINR must exceed a threshold $\gamma$, i.e., ${\rm SINR}_{j, \, i} > \gamma$ at node $j$ \cite{9187874}; otherwise, a transmission outage occurs. Meanwhile, we define two different node sets as
\begin{subequations} \label{S5-EQ28}
	\begin{align}
	\mathcal{Y}_{i} (t + 1) &\triangleq \left\{ j \left| \mathbb{I} \left( {\rm SINR}_{j, \, i} > \gamma \right) = 1, \, \forall j \in \mathcal{I} \setminus i \right. \right\},  \label{S5-EQ28a} \\
	\bar{\mathcal{I}} &\triangleq \{ i \in \mathcal{I} | \mathbb{I} \left( \mathcal{Y}_{i} (t + 1) \neq \emptyset \right) = 1 \}, \, \forall i \in \mathcal{I}, \label{S5-EQ28b}
	\end{align}
\end{subequations}
where $\mathcal{Y}_{i} (t + 1)$ represents the set of nodes that receive parameters from node $i$, and $\bar{\mathcal{I}}$ is the set of nodes that have been scheduled in this wireless network. More specifically, $\mathbb{I} \left( {\rm SINR}_{j, \, i} > \gamma \right)$ denotes a scheduled state whether node $i$ is scheduled by node $j$ while $\mathbb{I} \left( \mathcal{Y}_{i} (t + 1) \neq \emptyset \right)$ represents a scheduled state whether node $i$ has been scheduled at least once in this wireless network.

In theory, it is challenging to formulate an optimization problem about the SINR threshold $\gamma$ due to the varying $\Gamma_{W}$. In practice, as the value of $\gamma$ depends on the expectation of wireless channels and SINR, it can be optimized numerically by a grid search \cite{8870236}. To reduce the computational complexity, we consider two extreme cases for an appropriate $\gamma$. If the wireless network operates at a relatively high SINR, i.e., $\gamma \gg 0 $ \si{dB}, the number of scheduled nodes decreases monotonically, as implied by \eqref{S5-EQ28}. Hence, the number of waiting duration $\Gamma_{W}$ increases rapidly, yielding a large $\Gamma$. Clearly, a significant $\Gamma$ would deteriorate the convergence process and require more training iterations. In contrast, if the wireless network operates at low SINR, i.e., $\gamma \ll 0 \, {\rm dB}$, the number of scheduled nodes increases rapidly. However, the additional nodes exacerbate the competition for communication resources, resulting in a considerable transmission duration $\Gamma_{T}$, which also deteriorates the learning process. Consequently, we can conclude that an appropriate value $\gamma$ benefits making a balance between the waiting duration $\Gamma_{W}$ and the maximal transmission duration $\Gamma_{T}$ to minimize the total aggregated duration $\Gamma$. Moreover, it is widely known that SINR and $\gamma$ values depend heavily on the interference and channel fading among user nodes \cite{9562559}. As the waiting duration $\Gamma_{W}$ reserved for unscheduled nodes is fixed if the scheduling is determined by \eqref{S5-EQ28}, the maximal transmission duration $\Gamma_{T}$ can be employed to replace the optimal variable $\Gamma$ for efficient resource allocation. 

\subsection{Bandwidth Allocation}
As illustrated in Fig.~\ref{S4-Fig2}, $\Gamma_{j, \, i}$ denotes the transmission duration in each parameter transmission period. Then, $\Gamma_{T}$ can be computed as
\begin{equation} \label{S5-EQ26}
\Gamma_{T} = \max_{i \in \bar{\mathcal{I}}} \max_{j \in \mathcal{Y}_{i} (t + 1)} \Gamma_{j, \, i}.
\end{equation}
When the amount of transmission parameters is large, the parameter sparsity is necessarily accounted for to reduce transmission burden on wireless networks\footnote{It is well-known that when training a complex deep neural network model using stochastic gradient descent methods, model updates can be highly sparse. Indeed, it has been shown that when training some of the popular large-scale architectures, such as ResNet \cite{HeZRS16} or VGG \cite{SimonyanZ14a}, sparse levels of $\left[0.001, 0.01\right]$ provides a significant reduction in the communication load with almost no loss in their generalization performance \cite{WangniWLZ18}.}. In practice, each node only transmits non-zero parameters and their locations \cite{9562559}, such that $\Gamma_{j, \, i}$ can be computed as
\begin{equation} \label{S5-EQ27}
\Gamma_{j, \, i} = \dfrac{q_{i} S}{ T B_{i} \log_{2} \left( 1 + {\rm SINR}_{j, \, i} \right) }, \, \forall j \in \mathcal{Y}_{i} (t + 1), \, i \in \bar{\mathcal{I}}, 
\end{equation}
where $q_{i}$ is a sparsity ratio, $S$ is the total amount of the quantized parameter bits, $T$ is an average training latency for each iteration, and $B_{i}$ is an optimal bandwidth for node~$i$. In this paper, we set $q_{i} = 1$ since a weight decay $1 / 2 \|\bm{w}_{i}\|_{2}^{2}$ is considered for regularization terms.

\begin{algorithm}[t!] 
	\small
	\caption{The distributed FL algorithm with asynchronous parameter sharing at node $i$.}
	\label{S5-A1}
	\renewcommand{\algorithmicrequire}{\textbf{Input:}}
	\renewcommand{\algorithmicensure}{\textbf{Output:}}
	\begin{algorithmic}[1]
		\REQUIRE Setting wireless parameters similar to \cite{9435350}, learning error $\varepsilon$, $\{ \alpha_{j} \}_{j \in \mathcal{I} \setminus i}$, and $\mathcal{I}$. 
		\ENSURE The learning solution $\bm{w}_{i} (t)$.
		\STATE Initialize $t = t_{\rm S} = 0$, $\{ \bm{w}_{j} ( 0 ) \}_{j \in \mathcal{I}}$, $\mathcal{Y}_{i} (0) = \mathcal{I} \setminus i$, all states as $1$, $\{ {\rm SINR}_{j, \, i} \}_{j \in \mathcal{I} \setminus i}$, and $\{ R_{j} \}_{j \in \mathcal{I} \setminus i}$;
		\REPEAT
		\STATE Update $\bm{v}_{i} (t)$ and $\bm{w}_{i} (t + 1)$ as per \eqref{S4-EQ4b} and \eqref{S4-EQ5}, respectively; 
		\FOR {$j \in \mathcal{I} \setminus i$}
		\STATE Receive ${\rm SINR}_{j, \, i}$, the data rate $R_{j}$, the node state $Y_{j}$, the received state $J_{j, \, i}$, and the scheduled state $Q_{i, \, j}$ from node $j$;
		\IF {$Q_{i, \, j}=1$}
		\STATE Receive $\bm{w}_{j} ( \tau_{i, \, j} ( t ) )$ from node $j$ ($j \rightarrow i$), and feed a received state $J_{i, \, j} = 1$ back to node $j$; otherwise, feed $J_{i, \, j} = 0$ back.
		\ENDIF
		\ENDFOR
		\begin{framed}
			\IF {$\sum_{j \in \mathcal{Y}_{i} (t_{\rm S})} (1 - J_{j, \, i}) = 0$}
			\STATE Update the node set $\mathcal{Y}_{i} (t + 1)$ as per \eqref{S5-EQ28a} and $\bar{\mathcal{I}}$ as per \eqref{S5-EQ28b}, and $t_{\rm S} = t + 1$;
			\STATE Transmit $Q_{j, \, i}$ and $Y_{i} = \mathbb{I} \left( \mathcal{Y}_{i} (t + 1) \neq \emptyset \right)$ to other nodes  $\mathcal{I} \setminus i$. Specifically, $Q_{j, \, i} = 1$ if $j \in \mathcal{Y}_{i} (t + 1)$, and $Q_{j, \, i} = 0$ otherwise;
			\STATE Compute $R_{i} = \min_{j \in \mathcal{Y}_{i} (t + 1)} \log_{2} \left( 1 + {\rm SINR}_{j, \, i} \right)$, and transmit $R_{i}$ to other nodes $\mathcal{I} \setminus i$;
			\STATE Update $\{ B_{i} \}_{i \in \mathcal{I}}$ as per \eqref{S5-P1-a};
			\ENDIF
		\end{framed}
		\FOR {$j \in \mathcal{I} \setminus i$}
		\STATE Compute ${\rm SINR}_{i, \, j}$ like \eqref{S5-EQ22}, and transmit it to node $j$;
		\IF {$Q_{j, \, i} = 1$}
		\STATE Transmit $\bm{w}_{i} ( t + 1 )$ to node $j$ $(i \rightarrow j)$;
		\ENDIF		
		\ENDFOR
		\STATE $ t = t + 1 $;
		\UNTIL{$F_{i} ( \bm{v}_{i} (t) ) \leq \varepsilon$};
	\end{algorithmic}
\end{algorithm}

\begin{figure*}[t!]
	\centering
	\subfloat[A block diagram of Algorithm~\ref{S5-A1}.]{\includegraphics[width=0.8\linewidth]{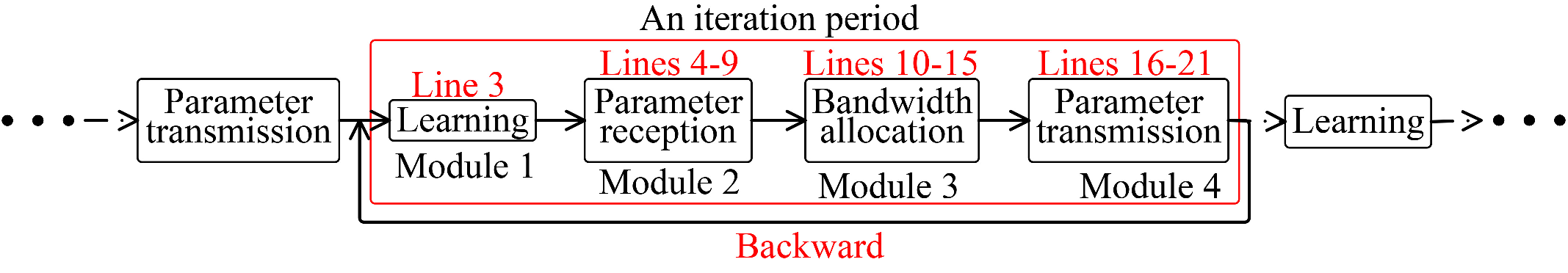}\label{S5-Fig4a}}
	\hfil
	\subfloat[Parameter Rx module.]{\includegraphics[width=0.23\linewidth]{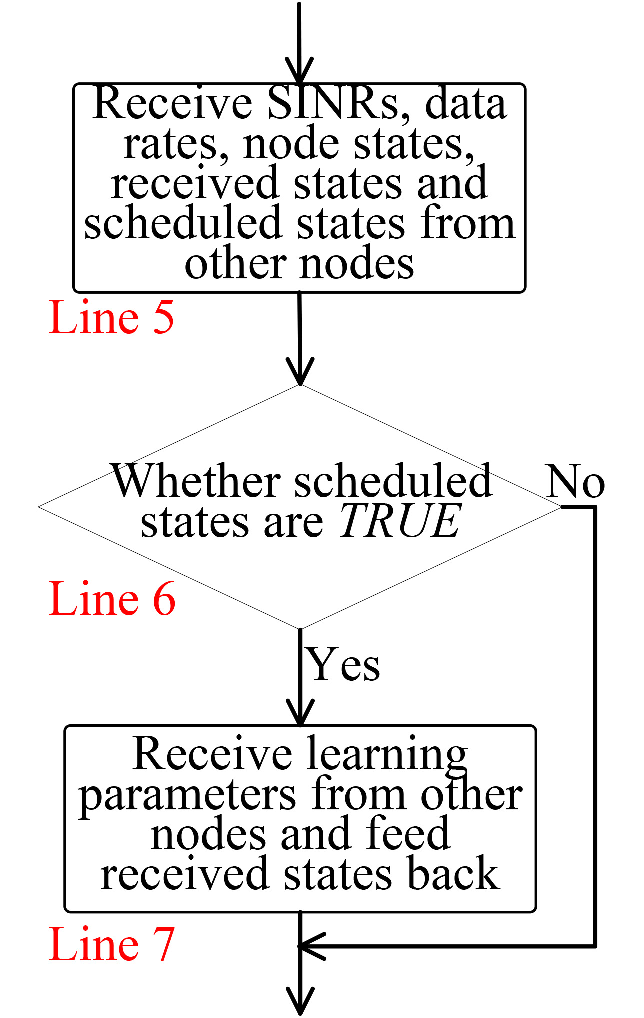}\label{S5-Fig4b}}
	\hfil
	\subfloat[Bandwidth allocation module.]{\includegraphics[width=0.23\linewidth]{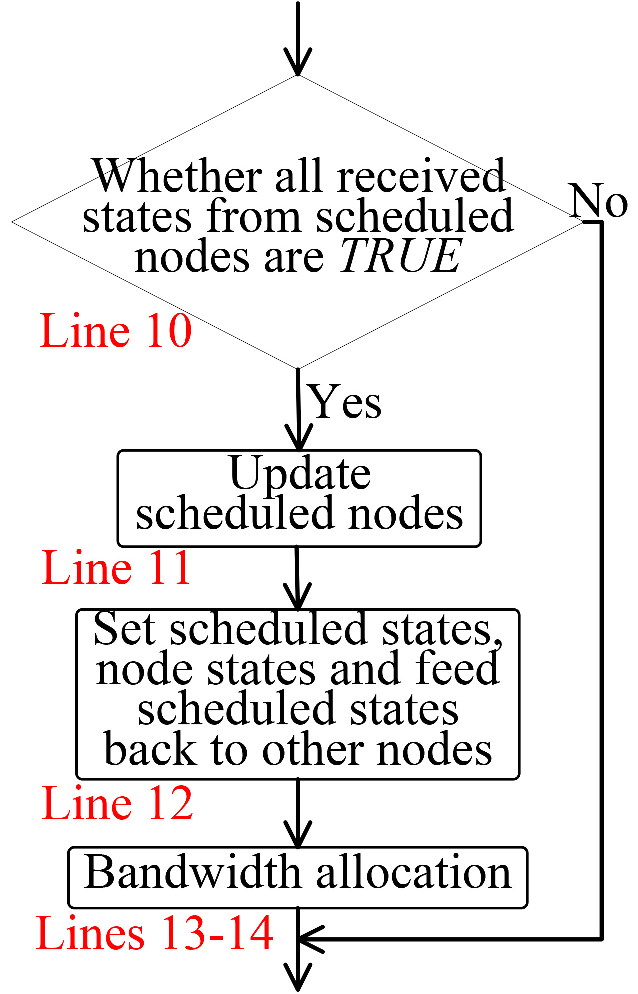}\label{S5-Fig4c}}
	\hfil
	\subfloat[Parameter Tx module.]{\includegraphics[width=0.23\linewidth]{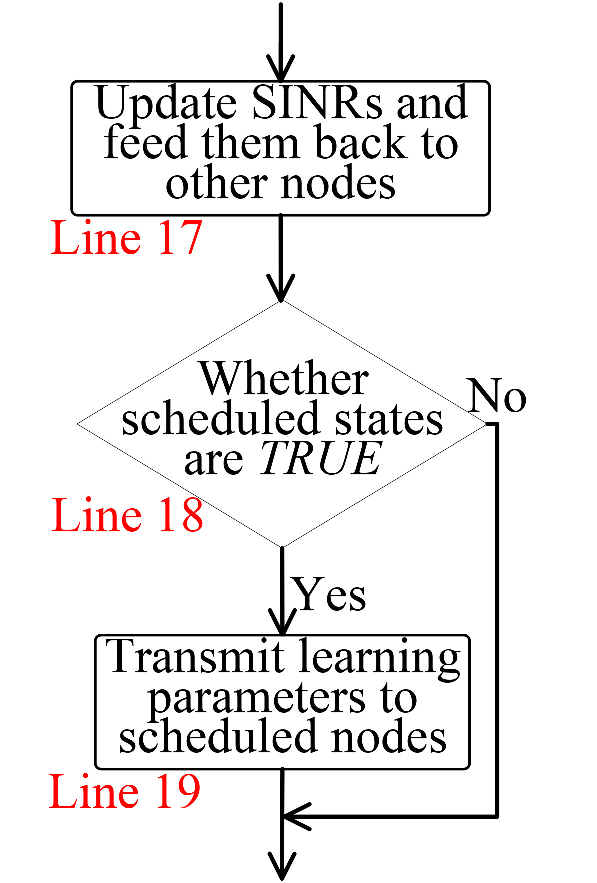}\label{S5-Fig4d}}
	\hfil
	\caption{The block diagram and built-in modules of Algorithm~\ref{S5-A1}.}
	\label{S5-Fig4}

\end{figure*}

To determine the transmission duration $\Gamma_{j, \, i}$ shown in \eqref{S5-EQ27}, we first assume that each node has a deterministic training latency $T$ at each node\footnote{In this paper, we assume that each node has the similar computation capability, and all data samples at each node have the same data size. Therefore, we set the average training latency $T$ as a constant for all nodes. A similar assumption is widely adopted in the literature, like \cite{9745059}.}, and then estimate $T$ in GPU modes. Unlike the serial mode of the CPU, the GPU processes datasets in parallel. When the training batch size is small, GPU can directly process the dataset simultaneously, yielding a constant training latency~$T_{0}$. On the contrary, the training latency $T$ grows linearly if the training batch size exceeds a threshold~$D_{\rm th}$. With this consideration, Assumption~$1$ in \cite{9252924} describes a computation model on the local training latency $T$ and the training batch size. It reveals that the local training latency grows linearly when the training batch size exceeds a threshold $D_{\rm th}$. Moreover, one training duration per iteration is usually as long as several seconds. It is because of the high computational complexity and the limited computation resource of nodes for processing large training sampling per iteration. However, communication time slots are relatively short \cite{9562559}. Therefore, the transmission duration $\Gamma_{j, \, i}$ computed by \eqref{S5-EQ27} is small. In light of \eqref{S5-EQ26} and \eqref{S5-EQ27}, we formulate an achievable transmission rate maximization problem as
\begin{subequations}
	\begin{align}
	\mathcal{P}{3} : \max_{ \left\{ B_{i} \right\}_{i \in \bar{\mathcal{I}}} } \min_{\substack{i \in \bar{\mathcal{I}}, \\ j \in \mathcal{Y}_{i} (t + 1)}} \, & B_{i} \log_{2} \left( 1 + {\rm SINR}_{j, \, i} \right) \\
	{\rm s.t.} \quad & \sum_{ i \in \bar{\mathcal{I}} } B_{i} \leq B, \\
	& B_{i} \geq 0, \, \forall i \in \bar{\mathcal{I}},
	\end{align}
\end{subequations}	
where $B$ is the total bandwidth. Clearly, $\mathcal{P}{3}$ is indeed a transmission delay minimization problem, which is convex and can be easily solved by the Karush-Kuhn-Tucker (KKT) conditions \cite[p. 243]{Convex2004Boyd}. Therefore, the optimal bandwidth allocation can be formalized in the following proposition.

With Proposition~\ref{S5-P1}, our decentralized FL model can be readily deployed in wireless distributed networks. Accordingly, we formalize a distributed algorithm in Algorithm~\ref{S5-A1}, where three types of states control the learning process. In particular, when the sharing parameter is transmitted from node $j$ to  $i$, the received state $J_{i, \, j}$ denotes whether node $i$ has received the parameter from node $j$ or not. The scheduled state $Q_{i, \, j}$ indicates whether node $i$ schedules node $j$ or not, while the state $Y_{i}$ reflects whether other nodes schedule node $i$ or not, i.e., $\mathcal{Y}_{i} (t + 1) \neq \emptyset$, which is used to update the node set $\bar{\mathcal{I}}$ defined by \eqref{S5-EQ28b}. 
\begin{proposition}[Bandwidth allocation] \label{S5-P1}
	Given the node set $\bar{\mathcal{I}}$ in the $(t + 1)^{\rm th}$ parameter transmission period, the optimal bandwidth allocation is given by 
	\begin{equation} \label{S5-P1-a}
	B_{i}^{*} = \left\{
	\begin{array}{rl}
	\dfrac{B}{R_{i}} \left( \sum\limits_{i^{\prime} \in \bar{\mathcal{I}}} \dfrac{1}{R_{i^{\prime}}} \right)^{-1}, & \text{if}~ i \in \bar{\mathcal{I}}; \\
	0, & \text{otherwise},
	\end{array}\right.
	\end{equation}
	where $R_{i} = \min_{j \in \mathcal{Y}_{i} (t + 1)} \log_{2} \left( 1 + {\rm SINR}_{j, \, i} \right), \, i \in \bar{\mathcal{I}}$, is an average uplink data rate during the $i^{\rm th}$ transmission period.
\end{proposition}

For better clarity of Algorithm~\ref{S5-A1}, Fig.~\ref{S5-Fig4a} shows the four built-in modules in each iteration period: a learning module, a parameter reception module, a bandwidth allocation module, and a parameter transmission module. In particular, Module 1, i.e., Line~$3$ of Algorithm~\ref{S5-A1}, states that the learning module includes local parameter aggregation and parameter updates. On the other hand, Figs.~\ref{S5-Fig4b}-\ref{S5-Fig4d} plots the flowcharts of Modules~2-4, respectively. In the parameter reception module, Fig.~\ref{S5-Fig4b} shows that the current node receives the SINR, data rate, three states above, and learning parameters from other nodes, which are realized in lines~4-9 of Algorithm~\ref{S5-A1}. Lines~$10$-$15$ realize the bandwidth allocation module shown in Fig.~\ref{S5-Fig4c}. Among them, line~10 means that node $i$ cannot transmit the current learning parameter to other nodes until it has finished all transmissions in the last parameter transmission process; line~11 sets a scheduling strategy for scheduled nodes. Each aggregated period has a separate iteration index $t_{\rm S}$ that records the moment of the parameter transmission. Line~12 feeds main states back to other nodes, and lines~13-14 compute the data rate and optimal bandwidth for the next iteration period. Lines~16-21 realize the parameter transmission module shown in Fig.~\ref{S5-Fig4d}, including two main steps: computing the SINR in line~17 and transmitting the learning parameters in lines~18-20.

\section{Simulation Results and Discussions} \label{Section-Simulation}
In this section, we present and discuss Monte-Carlo simulation results to exhibit the learning performance of the developed algorithm in terms of convergence analysis, node scheduling, and bandwidth allocation. In the pertaining simulation experiments, the workstation is equipped with an Intel(X) E$5$-$2620$ CPU and four NVIDIA GeForce RTX $3090$ graphic cards. We build a TensorFlow (TF) simulation environment and then train our proposed model by the deterministic gradient descent algorithm.

\subsection{Convergence Analysis}
This subsection evaluates the learning performance of Algorithm~\ref{S5-A1} without bandwidth allocation. For comparison purposes, we perform experiments on two typical datasets using CNN as follows: the MNIST \cite{MNIST2012Deng} and CIFAR datasets \cite{Learn2009Kriz}. In particular, the MNIST dataset contains $70,000$ $28 \times 28$ grayscale images of handwritten digits (more specifically, $60,000$ for training and $10,000$ for testing). The CIFAR dataset contains $60,000$ $32 \times 32$ color images (more specifically,  $50,000$ for training and $10,000$ for testing). The images in each dataset are divided into $10$ classes with an equal number of images per class. The CNN has $9$ layers with the following structure: $5 \times 5 \times 32$ Convolutional $\rightarrow$ $2 \times 2$ MaxPool $\rightarrow$ Local Response Normalization $\rightarrow$ $5 \times 5 \times 32$ Convolutional $\rightarrow$ Local Response Normalization $\rightarrow$ $2 \times 2$ MaxPool $\rightarrow$ $z \times 256$ Fully connected $\rightarrow$ $256 \times 10$ Fully connected $\rightarrow$ Softmax, where $z = 1568$ for the MNIST and $z = 2048$ for the CIFAR \cite{8664630}. In the simulation experiments, we consider five typical schemes: the centralized FL averaging model (FedAv for short) \cite{ZhangSFYA21}; adaptive FL model (AdaFL for short) \cite{8664630}; the FedAv-partial model (FedAP for short), in which the parameter server selects partial nodes for the global updating to avoid the impacts of stragglers \cite{9562538}; the semi-asynchronous FL model (FedSA for short), where the parameter server aggregates a certain number of local models by their arrival order in each round \cite{9562538}; and Algorithm~\ref{S5-A1} without bandwidth allocation, i.e., the pseudo-code in the box of Algorithm~\ref{S5-A1} is excluded (Alg.~$1$ w/o BWA for short). Note that Algorithm~\ref{S5-A1} is reduced for a fair comparison. 

\begin{figure*}[t!]
	\centering
	\subfloat[Training loss in the MNIST dataset.]{\includegraphics[width=0.45\linewidth]{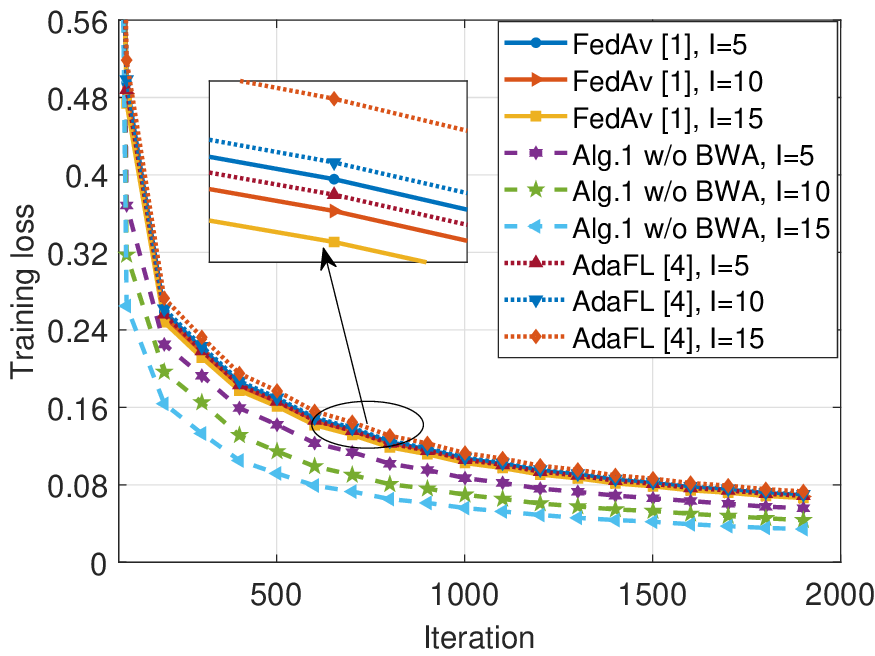} \label{Fig5a}}
	\hfil
	\subfloat[Testing accuracy in the MNIST dataset.]{\includegraphics[width=0.45\linewidth]{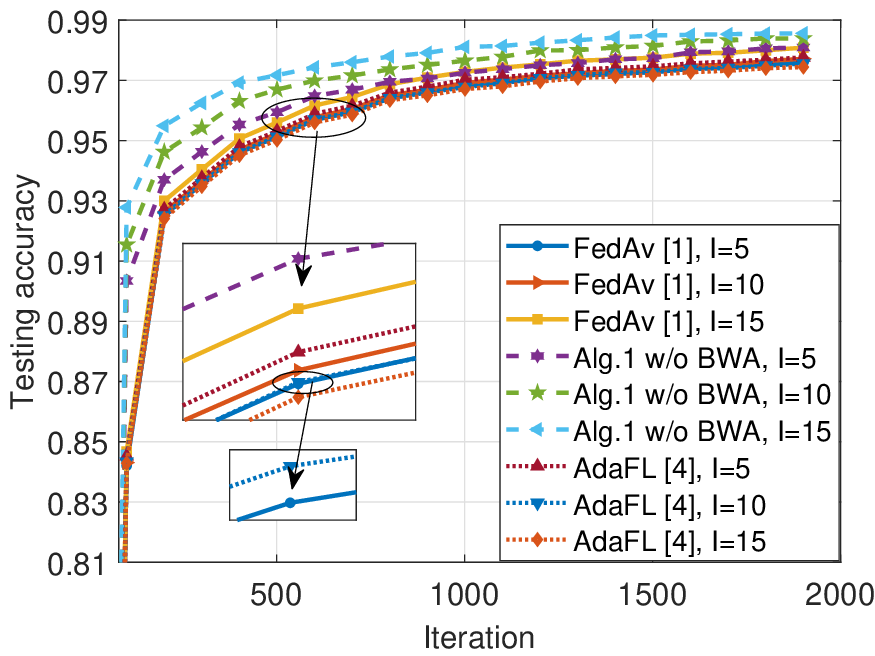} \label{Fig5b}}
	\hfil
	\subfloat[Training loss in the CIFAR dataset.]{\includegraphics[width=0.45\linewidth]{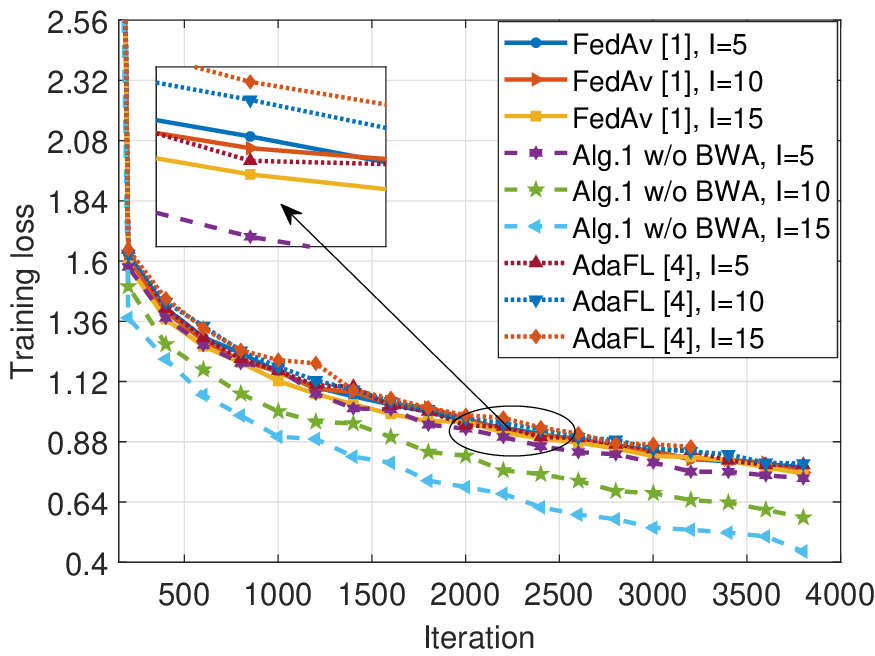} \label{Fig5c}}
	\hfil
	\subfloat[Testing accuracy in the CIFAR dataset.]{\includegraphics[width=0.45\linewidth]{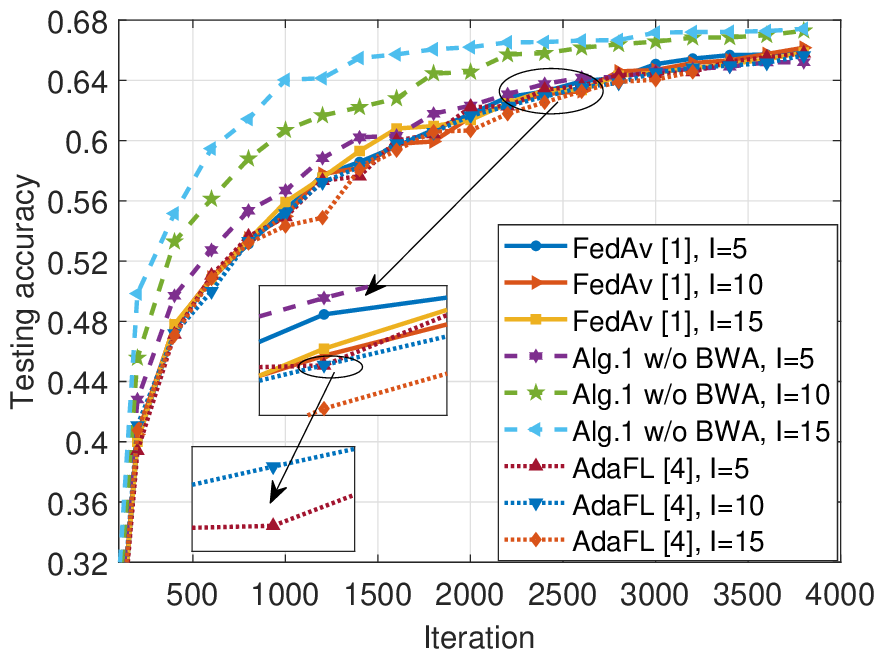} \label{Fig5d}}
	\hfil			
	\caption{Learning performance vs. the number of iterations, where $\Gamma = 5$ and $\eta = 0.016$ are set for the MNIST dataset whereas $\eta = 0.02$ is set for the CIFAR dataset.}
	\label{Fig5}
\end{figure*}

Figure~\ref{Fig5} depicts the learning performance versus the number of iterations, where the number of nodes is set to $I = 5, 10, 15$. On the one hand, we observe from Fig.~\ref{Fig5a} that the training losses of Algorithm~\ref{S5-A1} are lower than those for comparison. The reason behind this observation is that our decentralized FL model of $\mathcal{P}1$ has a less-nonconvex objective function than the centralized FL averaging model \cite{ZhangSFYA21} and adaptive FL model \cite{8664630}. Also, it is observed that when the number of nodes increases from $5$ to $10$ till $15$, the training loss of Algorithm~\ref{S5-A1} converges to a lower and lower value. This observation is because the normalized duality gap decreases as the number of nodes increases. Thus, our proposed model becomes more and more convex when a large dataset is distributed over each node uniformly. Consequently, this makes the learning process faster to obtain an optimal solution. On the other hand, Fig.~\ref{Fig5b} also shows that the testing accuracies of Algorithm~\ref{S5-A1} outperform the benchmark algorithms under study, and more nodes involved lead to higher testing accuracy, as expected. As far as the CIFAR dataset is concerned, Figs.~\ref{Fig5c}-\ref{Fig5d} show similar performance to Figs.~\ref{Fig5a}-\ref{Fig5b}, respectively. Accordingly, in the rest of this section, we only report the results regarding the MNIST dataset to reduce redundancy.

\begin{figure*}[t!]
	\centering
	\subfloat[Training loss.]{\includegraphics[width=0.45\linewidth]{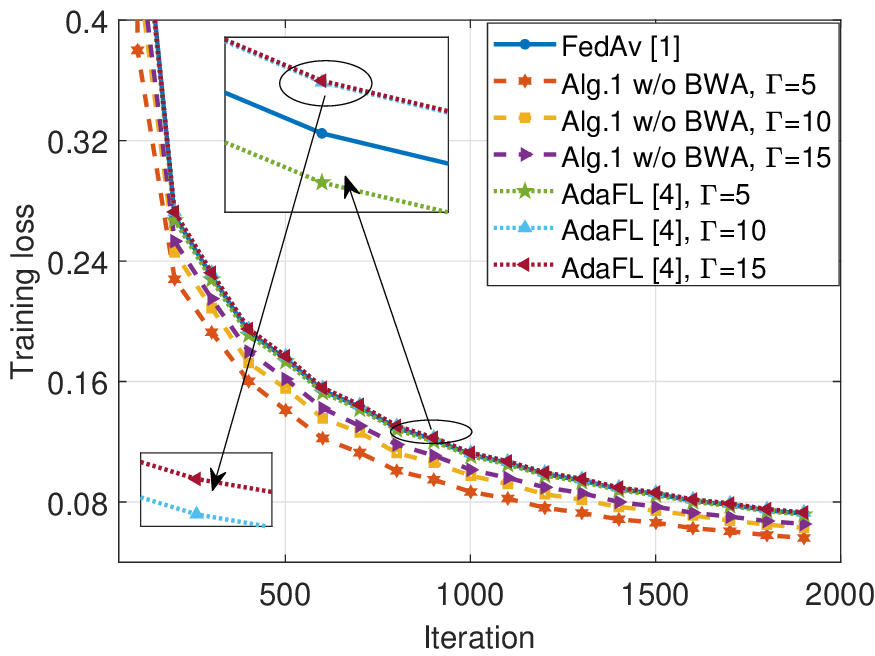} \label{Fig6a}}
	\hfil
	\subfloat[Testing accuracy.]{\includegraphics[width=0.45\linewidth]{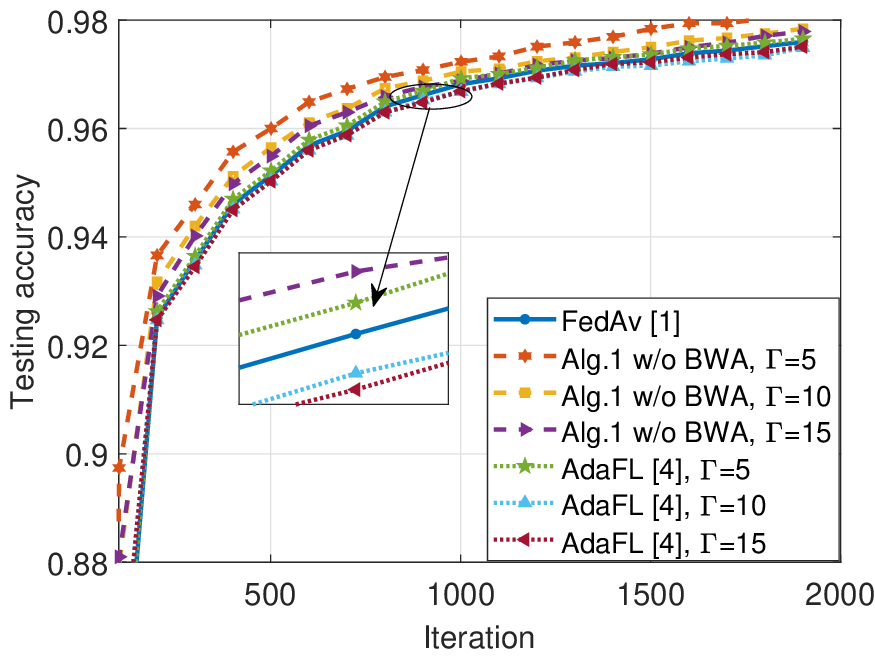} \label{Fig6b}}
	\hfil		
	\caption{Learning performance vs. the number of iterations, where $I = 5$ and $\eta = 0.016$ are set for the MNIST dataset.}
	\label{Fig6}
	\vspace{-5pt}
\end{figure*}

Figure~\ref{Fig6} illustrates the learning performance versus the number of iterations, where the transmission delays are set to $\Gamma = 5, 10, 15$. In particular, the centralized FL averaging model is independent of $\Gamma$ as the transmission delays and aggregated durations do not influence learning performance. On the one hand, Fig.~\ref{Fig6a} shows that the training losses of Algorithm~\ref{S5-A1} are lower than the benchmark algorithms under study. Moreover, as $\Gamma$ decreases from $15$ to $10$ till $5$, the training loss of Algorithm~\ref{S5-A1} becomes smaller and smaller. The reason behind these observations is that, given a smaller transmission delay $\Gamma$, our decentralized FL model can faster aggregate sharing parameters from other nodes, thus leading to a lower upper bound and a faster convergence rate, by Theorem~\ref{S5-T2}. Instead, when $\Gamma$ becomes large, the proposed model can only obtain sharing parameters with a large delay, slightly slowing down the convergence. On the other hand, Fig.~\ref{Fig6b} shows that the testing accuracy of Algorithm~\ref{S5-A1} outperforms the benchmark algorithms under study, regardless of $\Gamma$. 

\subsection{Effect of Node Scheduling and Bandwidth Allocation}
This subsection illustrates the effectiveness of node scheduling and bandwidth allocation. In the pertaining simulation experiments, we adopt similar wireless parameters as those reported in \cite{9435350}, and their default values are summarized in Table~\ref{S5-T3} unless specified otherwise. Apart from the conventional centralized FL averaging model (FedAv for short) \cite{ZhangSFYA21} and Algorithm~\ref{S5-A1} with optimal resource allocation (Alg.~$1$ for short), we also consider two other benchmark ones: the distributed FL algorithms with uniform bandwidth allocation (Distributed+Uniform for short) or random bandwidth allocation (Distributed+Random for short). 

\begin{table}[t!] 
	\small
	\centering
	\caption{Wireless parameters \cite{9435350}.}
	\begin{tabular}{!{\vrule width 1.2pt}c !{\vrule width 0.8pt} c !{\vrule width 1.5pt}}
		\Xhline{1.2pt}
		\textbf{Parameter} & \textbf{Value} \\
		\Xhline{1.0pt}
		Cell radius & $500$ \si{m} \\
		\hline
		Path loss exponent ($\alpha$) & $4$ \\
		\hline
		Bandwidth ($B$) & $10$ \si{MHz} \\
		\hline
		Transmit power ($P$) & $30$ \si{dBm} \\
		\hline
		Noise power density & $-174$ \si{dBm/Hz} \\
		\hline
		\# of quantization bits per sample & $16$ \si{bits} \\
		\Xhline{1.0pt}
	\end{tabular}
	\label{S5-T3}
	\vspace{-10pt}
\end{table}

\begin{figure*}[t!]
	\centering
	\subfloat[Training loss.]{\includegraphics[width=0.45\linewidth]{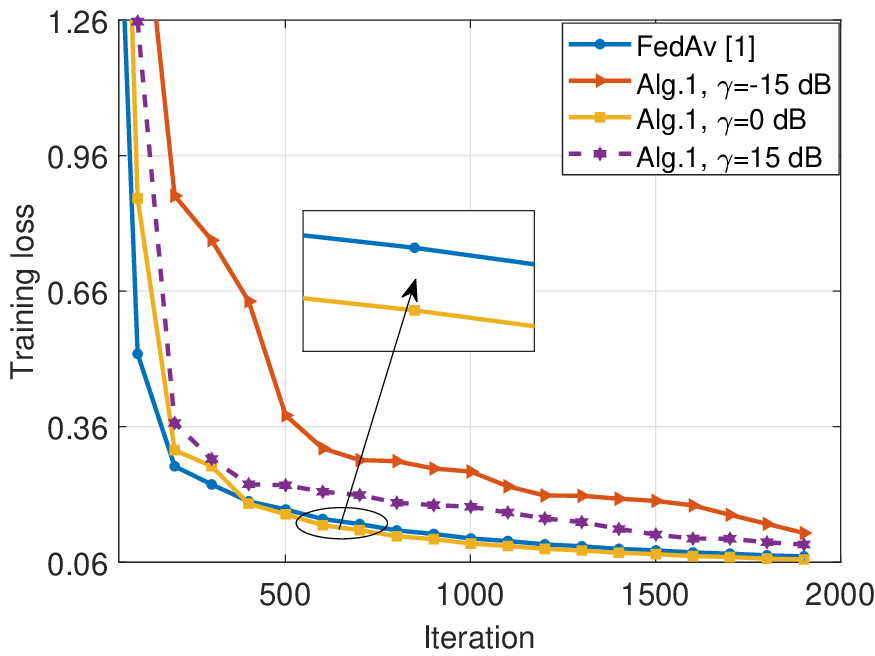} \label{Fig7a}}
	\hfil
	\subfloat[Testing accuracy.]{\includegraphics[width=0.45\linewidth]{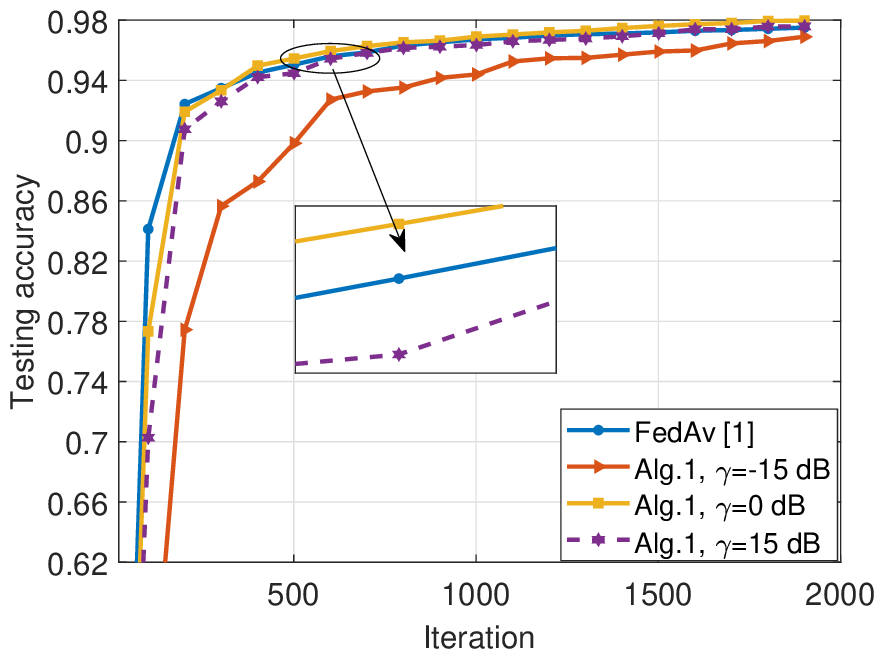} \label{Fig7b}}
	\hfil	
	\caption{Learning performance vs. the number of iterations, where $I = 5$, $B = 10$ \si{MHz}, and $\eta = 0.016$ are set for the MNIST dataset.}
	\label{Fig7}
\end{figure*}

Figure~\ref{Fig7} illustrates the effect of node scheduling by adjusting different SINR thresholds (say, $\gamma$) in Algorithm~\ref{S5-A1}. On the one hand, Fig.~\ref{Fig7a} shows that, given $\gamma = 0$ \si{dB}, the training loss of Algorithm~\ref{S5-A1} is lower than those of $\gamma = -15$ \si{dB} and $\gamma = 15$ \si{dB}, and perform a bit smaller than the centralized algorithm. These observations are because, given a low SINR threshold $\gamma$, many nodes are scheduled, and the channels with poor SINR cannot obtain sufficient bandwidth. As a result, the transmission delay and aggregated duration become large, thus leading to a low convergence rate. Meanwhile, many nodes are not scheduled given a high SINR threshold $\gamma$. Hence, the waiting duration $\Gamma_{W}$ reserved for unscheduled nodes increases, resulting in poor learning performance. On the other hand, Fig.~\ref{Fig7b} shows that the testing accuracy with $\gamma = 0$ \si{dB} is higher than those of $\gamma = -15$ \si{dB} and $\gamma = 15$ \si{dB}. Therefore, adopting a medium SINR threshold is critical to achieving good convergence performance of our decentralized FL model.

\begin{figure*}[t!]
	\centering
	\subfloat[Training loss.]{\includegraphics[width=0.45\linewidth]{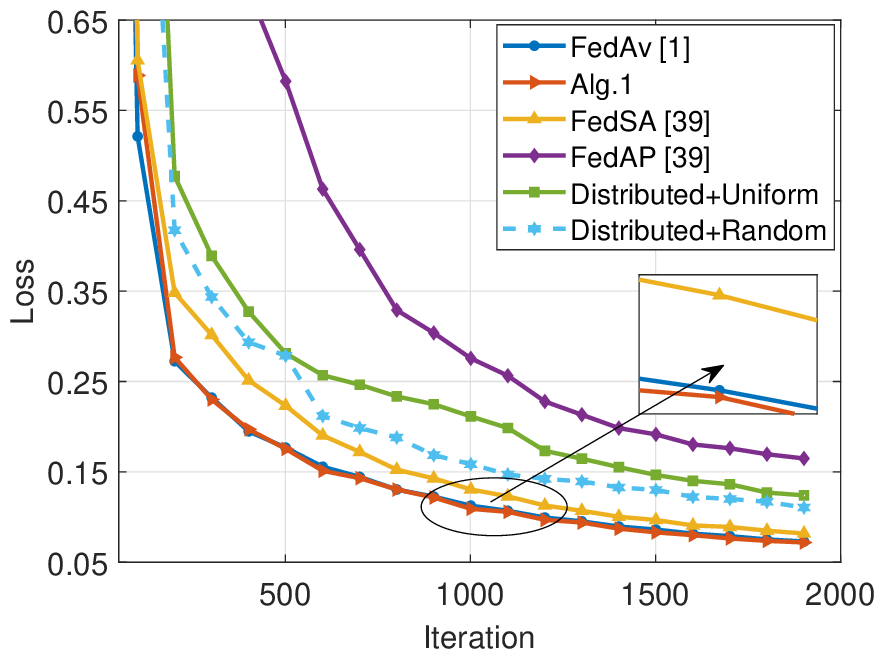} \label{Fig8a}}
	\hfil
	\subfloat[Testing accuracy.]{\includegraphics[width=0.45\linewidth]{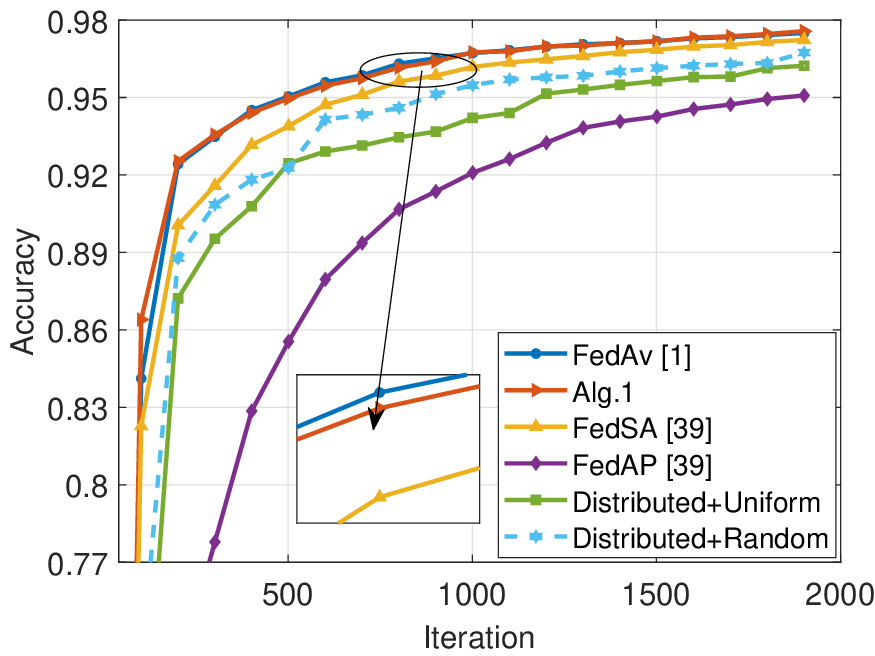} \label{Fig8b}}
	\hfil	
	\caption{Learning performance of different learning algorithms, where $I = 5$, $\gamma = 0$ \si{dB}, $B = 10$ \si{MHz}, and $\eta = 0.016$ are set for the MNIST dataset.}
	\label{Fig8}
	\vspace{-10pt}
\end{figure*}

Figure~\ref{Fig8} compares Algorithm~\ref{S5-A1} with state-of-the-art algorithms in resource-limited settings, including FedSA and FedAP, and Algorithm~\ref{S5-A1} with different bandwidth allocation mechanisms, including uniform and random allocation mechanisms. On the one hand, Fig.~\ref{Fig8a} shows that Algorithm~\ref{S5-A1} and the centralized algorithm achieve similar training losses, both lower than the benchmark ones under study. The reason stems from the fact that Algorithm~\ref{S5-A1} can allocate bandwidth optimally such that the transmission delay and aggregated duration can also be minimized. Theorem~\ref{S5-T2} implies that a low aggregated duration leads to a low convergence bound and accelerates the training process. However, the random and uniform bandwidth allocations are not optimal, thus resulting in a longer transmission delay and aggregated duration than Algorithm~\ref{S5-A1}. Moreover, it is observed that Algorithm~\ref{S5-A1} and FedSA have a lower learning loss than FedAP. The reason is that the FedAP algorithm only schedules partial nodes for the global aggregation and has data deficiency due to the communication stragglers. Algorithm~\ref{S5-A1} also has a lower learning loss than FedSA due to its sharing parameters and less non-convexity. On the other hand, Fig.~\ref{Fig8b} shows that the testing accuracy of Algorithm~\ref{S5-A1} is also higher than the others. Therefore, we can infer that the decentralized FL model with optimal bandwidth allocation enables efficient learning even in resource-limited wireless networks.

\section{Concluding Remarks} \label{Section-Conclusions}
This paper has developed a decentralized FL model with less non-convex summation functions to adapt to distributed network architecture, especially for resource-constrained large-scale IoT networks. With the increasing number of nodes, this paper has derived a duality gap to control the non-convexity of our proposed decentralized FL model and improve the learning performance. To realize efficient learning, this paper has also proposed a gradient descent algorithm with asynchronous parameter sharing and a convergence analysis to mitigate the influence of stochastic transmission delays. Moreover, we established the relationship between the convergence bound and the transmission delay for efficient communication and excellent learning performance in resource-constrained IoT networks. It improved the learning performance by minimizing the transmission delay. In future works, we will incorporate the data distribution divergence over nodes and design a new gradient descent to deal with insufficient data samples.

\appendices
\numberwithin{equation}{section}

\section{Proof of Theorem~\ref{S3-T1}} \label{SA-A}	
We start to define two subsets:
\begin{subequations}
	\begin{align}
	\mathcal{Y}_{i} &{} \triangleq \left\{ \bm{y}_{i} \left| \bm{y}_{i} = \left[ r ( \bm{w}_{i} ), \, F_{i} \left( ( 1 - \alpha_{i} ) \bm{w}_{i|{\rm sh}} + \alpha_{i} \bm{w}_{i} \right) \right]^{T} \right. \right\}, \nonumber \\
	\mathcal{Y} &{} \triangleq \sum_{i = 1}^{I} \alpha_{i} \mathcal{Y}_{i} = \left\{ \bm{y} \left| \bm{y} = \sum_{i = 1}^{I} \alpha_{i} \bm{y}_{i}, \right. \, \forall \bm{y}_{i} \in \mathcal{Y}_{i}, \, i \in \mathcal{I} \right\}. \nonumber
	\end{align}
\end{subequations}
From the geometric properties of duality \cite[pp. 232-235]{Convex2004Boyd}, we can obtain 
\begin{subequations}
	\begin{align}
	\inf (\mathcal{P}1) &{} = \inf \{ f \left| (r, \, f) \in \mathcal{Y} \right. \} , \\
	\sup (\mathcal{P}2) &{} = \inf \{ f \left| (r, \, f) \in { \rm conv } ( \mathcal{Y} ) \right. \} ,
	\end{align}
\end{subequations}
where ${\rm conv} ( \mathcal{Y} )$ means the convex hull of $\mathcal{Y}$.

Now, we use the Shapley-Folkman lemma \cite{BiT20} to prove Theorem~\ref{S3-T1}. Let $( \bar{r}, \, \sup ( \mathcal{P}2 ) ) \in {\rm conv} ( \mathcal{Y} )$, by the Shapley-Folkman lemma, there is a subset $\bar{ \mathcal{I} } \subseteq \mathcal{I}$ of $\left| \bar{\mathcal{I}} \right| = 2$ so that 
\begin{align}
\bar{r}& {}= \sum_{ i \in \mathcal{I} \backslash \bar{ \mathcal{I} } } \alpha_{i} r ( \bar{\bm{w}}_{i} ) + \sum_{ i \in \bar{ \mathcal{I} } } \alpha_{i} \bar{r}_{i}, \label{SA-EQ-A3a} \\
\sup (\mathcal{P}2)& {}= \sum_{ i \in \mathcal{I} \backslash \bar{ \mathcal{I} } } \alpha_{i} F_{i} \left( ( 1 - \alpha_{i} ) \bm{w}_{i|{\rm sh}} + \alpha_{i} \bar{\bm{w}}_{i} \right) + \sum_{ i \in \bar{ \mathcal{I} } } \alpha_{i} \bar{f}_{i}, \nonumber
\end{align}
where $( \bar{r}_{i}, \, \bar{f}_{i} ) \in {\rm conv} ( \mathcal{Y}_{i} ) , \forall i \in \bar{ \mathcal{I} }$ and $ \bar{\bm{w}}_{i} \in \mathcal{W}_{i}, \, \forall i \in \mathcal{I} \backslash \bar{ \mathcal{I} } $. As a result, we can represent the elements of ${\rm conv} (\mathcal{Y}_{i})$ for all $i \in \bar{\mathcal{I}}$ as follows
\begin{equation} \label{SA-EQ-A4}
\bar{r}_{i} = \sum_{j = 1}^{N_{i}} \theta_{j} r ( \bm{x}_{j} ) \geq r \left( \sum_{j = 1}^{N_{i}} \theta_{j} \bm{x}_{j} \right), 
\end{equation}
where $\sum_{j = 1}^{N_{i}} \theta_{j} = 1, \, \forall \theta_{j} \geq 0,  \, \bm{x}_{j} \in \mathcal{W}_{i}$. By definition of $\Delta_{i}$ shown in Theorem~\ref{S3-T1}, we can obtain 
\begin{align}
&\sup (\mathcal{P}2) = \sum_{j = 1}^{N_{i}} \theta_{j} F_{i} \left( ( 1 - \alpha_{i} ) \bm{w}_{i|{\rm sh}} + \alpha_{i} \bm{x}_{j} \right) \nonumber \\
&\geq \tilde{F}_{i} \left( \sum_{j = 1}^{N_{i}} \theta_{j} \bm{x}_{j} \right) \geq \hat{F}_{i} \left( \sum_{j = 1}^{N_{i}} \theta_{j} \bm{x}_{j} \right) - \Delta_{i}. \label{SA-EQ-A5}
\end{align}

In light of \eqref{SA-EQ-A3a}-\eqref{SA-EQ-A5}, we have
\begin{subequations}
	\begin{align}
	&{} \sum_{ i \in \mathcal{I} \backslash \bar{ \mathcal{I} } } \alpha_{i} r ( \bar{\bm{w}}_{i} ) + \sum_{ i \in \bar{ \mathcal{I} } } \alpha_{i} r \left( \sum_{j = 1}^{N_{i}} \theta_{j} \bm{x}_{j} \right) \leq \bar{r}, \label{SA-EQ-A6a} \\
	&{} \sum_{ i \in \mathcal{I} \backslash \bar{ \mathcal{I} } } \alpha_{i} F_{i} \left( ( 1 - \alpha_{i} ) \bm{w}_{i|{\rm sh}} + \alpha_{i} \bar{\bm{w}}_{i} \right) + \sum_{ i \in \bar{ \mathcal{I} } } \alpha_{i} \hat{F}_{i} \left( \sum_{j = 1}^{N_{i}} \theta_{j} \bm{x}_{j} \right) \nonumber \\
	&{} \leq \sup (\mathcal{P}2) + \sum_{i \in \bar{\mathcal{I}}} \alpha_{i} \Delta_{i}, \label{SA-EQ-A6b}
	\end{align}
\end{subequations}
then, we can find a $\bar{\bm{w}}_{i} \in \mathcal{W}_{i}$ for $i \in \bar{\mathcal{I}}$ such that $r \left( \bar{\bm{w}}_{i} \right) \leq r \left( \sum_{j = 1}^{N_{i}} \theta_{j} \bm{x}_{j} \right)$ and
\begin{equation} \label{SA-EQ-A7}
F_{i} \left( ( 1 - \alpha_{i} ) \bm{w}_{i|{\rm sh}} + \alpha_{i} \bar{\bm{w}}_{i} \right) = \hat{F}_{i} \left( \sum_{j = 1}^{N_{i}} \theta_{j} \bm{x}_{j} \right).
\end{equation}

In light of \eqref{SA-EQ-A6a}-\eqref{SA-EQ-A7}, it yields $\sum_{ i \in \mathcal{I} } \alpha_{i} r ( \bar{\bm{w}}_{i} ) \leq \bar{r}$, $\bar{r} \triangleq \sum_{i = 1}^{I} \alpha_{i} K_{i}$ and
\begin{align}
&\inf ( \mathcal{P}1 ) \leq \sum_{ i \in \mathcal{I} } \alpha_{i} F_{i} \left( ( 1 - \alpha_{i} ) \bm{w}_{i|{\rm sh}} + \alpha_{i} \bar{\bm{w}}_{i} \right) \nonumber\\
&\leq \sup (\mathcal{P}2) + \sum_{i \in \bar{\mathcal{I}}} \alpha_{i} \Delta_{i} \leq \sup (\mathcal{P}2) + 2 \alpha_{\max} \Delta_{\rm worst}. \label{EQ-SA-A10b}
\end{align}
The proof is complete by performing some algebraic manipulations of \eqref{EQ-SA-A10b}. 

\section{Proof of Theorem~\ref{S5-T2}} \label{SA-B}
For notational convenience, we define $\bm{s}_{i} (t) = \bm{0}$ if $t < 0$, $\bm{g} (t) \triangleq \left[ \bm{s}_{1}^{T} (t), \, \bm{s}_{2}^{T} (t), \, \cdots, \, \bm{s}_{I}^{T} (t) \right]^{T}$, and $\bar{\bm{s}} ( t ) \triangleq \sum_{ i = 1 }^{I} \bm{s}_{i}^{T} (t)$. By using the gradient descent algorithm, we obtain
\begin{subequations}
	\begin{align}
	\lefteqn{\bar{F} \left( \bm{w} ( t + 1 ) \right)} \nonumber \\
	&= \bar{F} \left( \bm{w} ( t ) + \eta \bar{\bm{s}} ( t ) \right) \quad \text{(Updating from \eqref{S4-EQ6})} \nonumber \\
	&\leq \bar{F} \left( \bm{w} ( t ) \right) + \eta \left\langle \nabla \bar{F} \left( \bm{w} ( t ) \right), \, \bar{\bm{s}} ( t ) \right\rangle + \dfrac{L_{1}}{2} \eta^{2} \| \bar{\bm{s}} ( t ) \|_{2}^{2} \nonumber \\
	& \quad \text{(From \eqref{S4-EQ19} of Lemma~\ref{S4-L1})} \nonumber \\
	&\leq \bar{F} \left( \bm{w} ( t ) \right) + \eta \sum_{i = 1}^{I} \left\langle \nabla \bar{F} \left( \bm{w} ( t ) \right), \, \bm{s}_{i} ( t ) \right\rangle + \dfrac{L_{1}}{2} I \eta^{2} \| \bm{g} ( t ) \|_{2}^{2} \nonumber \\
	& \quad \text{(The Jensen's inequality)} \nonumber \\
	&\leq \bar{F} \left( \bm{w} ( t ) \right) + \eta \sum_{i = 1}^{I} \left\langle \nabla \bar{F} \left( \bm{w} ( t ) \right) - \nabla \bar{F} \left( \bm{v}_{i} ( t ) \right), \, \bm{s}_{i} ( t ) \right\rangle \nonumber \\
	&\quad{} + \, \eta \sum_{i = 1}^{I} \left\langle \nabla \bar{F} \left( \bm{v}_{i} ( t ) \right), \, \bm{s}_{i} ( t ) \right\rangle + \dfrac{L_{1}}{2} I \eta^{2} \| \bm{g} ( t ) \|_{2}^{2} \nonumber \\
	& \quad \text{(Adding a zero term)} \nonumber \\
	&\leq \bar{F} \left( \bm{w} ( t ) \right) + \eta \sum_{i = 1}^{I} \left\langle \nabla \bar{F} \left( \bm{w} ( t ) \right) - \nabla \bar{F} \left( \bm{v}_{i} ( t ) \right), \, \bm{s}_{i} ( t ) \right\rangle \nonumber \\
	&\quad{} + \, \eta \sum_{i = 1}^{I} \sum_{ j = 1 }^{I} \left\langle \nabla_{j} F_{j} ( \bm{v}_{i} ( t ) ), \, \bm{s}_{i} ( t ) \right\rangle + \dfrac{L_{1}}{2} I \eta^{2} \| \bm{g} ( t ) \|_{2}^{2}  \nonumber \\
	&\quad \text{(From \eqref{S4-EQ13} of Lemma~\ref{S4-L1})} \nonumber \\
	&\leq \bar{F} \left( \bm{w} ( t ) \right) + \eta \sum_{i = 1}^{I} \sum_{ j = 1, j \neq i }^{I} \left\langle \nabla_{j} F_{j} ( \bm{v}_{i} ( t ) ), \, \bm{s}_{i} ( t ) \right\rangle \nonumber \\
	&\quad{} + \, \dfrac{L_{1}}{2} I \eta^{2} \| \bm{g} ( t ) \|_{2}^{2} + \eta \sum_{i = 1}^{I} \left\langle \nabla_{i} F_{i} ( \bm{v}_{i} ( t ) ), \, \bm{s}_{i} ( t ) \right\rangle \nonumber \\
	&\quad{} + \, \eta \sum_{i = 1}^{I} \left\langle \nabla \bar{F} \left( \bm{w} ( t ) \right) - \nabla \bar{F} \left( \bm{v}_{i} ( t ) \right), \, \bm{s}_{i} ( t ) \right\rangle \nonumber \\
	&\leq \bar{F} \left( \bm{w} ( t ) \right) + \eta \sum_{i = 1}^{I} \left\langle \nabla_{i} F_{i} ( \bm{v}_{i} ( t ) ), \, \bm{s}_{i} ( t ) \right\rangle + \dfrac{L_{1}}{2} I \eta^{2} \| \bm{g} ( t ) \|_{2}^{2} \nonumber \\
	&\quad{} + \, \eta \sum_{i = 1}^{I} \sum_{ j = 1, j \neq i }^{I} \left\langle \nabla_{j} F_{j} ( \bm{v}_{i} ( t ) ) - \nabla_{j} F_{j} ( \bm{v}_{j} ( t ) ), \, \bm{s}_{i} ( t ) \right\rangle \nonumber \\
	&\quad{} + \, \eta \sum_{i = 1}^{I} \left\langle \nabla \bar{F} \left( \bm{w} ( t ) \right) - \nabla \bar{F} \left( \bm{v}_{i} ( t ) \right), \, \bm{s}_{i} ( t ) \right\rangle \nonumber \\
	&\quad{} + \, \eta \sum_{i = 1}^{I} \sum_{ j = 1, j \neq i }^{I} \left\langle \nabla_{j} F_{j} ( \bm{v}_{j} ( t ) ), \, \bm{s}_{i} ( t ) \right\rangle \nonumber \\
	& \quad{} \text{(Adding a zero term)} \nonumber \\
	&\leq \bar{F} \left( \bm{w} ( t ) \right) + \eta I \sum_{i = 1}^{I} \left\langle \nabla_{i} F_{i} ( \bm{v}_{i} ( t ) ), \, \bm{s}_{i} ( t ) \right\rangle + \dfrac{L_{1}}{2} I \eta^{2} \| \bm{g} ( t ) \|_{2}^{2} \nonumber \\
	&\quad{} + \, \eta \sum_{i = 1}^{I} \sum_{ j = 1, j \neq i }^{I} \left\langle \nabla_{j} F_{j} ( \bm{v}_{j} ( t ) ), \, \bm{s}_{i} ( t ) - \bm{s}_{j} ( t ) \right\rangle \nonumber\\
	&\quad{} + \, \eta \sum_{i = 1}^{I} \sum_{ j = 1, j \neq i }^{I} \left\langle \nabla_{j} F_{j} ( \bm{v}_{i} ( t ) ) - \nabla_{j} F_{j} ( \bm{v}_{j} (t) ), \, \bm{s}_{i} ( t ) \right\rangle \nonumber \\
	&\quad{} + \, \eta \sum_{i = 1}^{I} \left\langle \nabla \bar{F} \left( \bm{w} (t) \right) - \nabla \bar{F} \left( \bm{v}_{i} (t) \right), \, \bm{s}_{i} (t) \right\rangle \nonumber \\
	&\leq \bar{F} \left( \bm{w} ( t ) \right) - \eta I \sum_{i = 1}^{I} \dfrac{1}{L_{3}} \| \bm{s}_{i} ( t ) \|_{2}^{2} + \dfrac{\eta \delta}{L_{2}} \sum_{i = 1}^{I} \sum_{ j = 1, j \neq i }^{I} \| \bm{s}_{j} ( t ) \|_{2}^{2} \nonumber \\
	&\quad{} + \, \eta \sum_{i = 1}^{I} L_{1} \| \bm{w} ( t ) - \bm{v}_{i} ( t ) \|_{2} \| \bm{s}_{i} ( t ) \|_{2} + \dfrac{L_{1}}{2} I \eta^{2} \| \bm{g} ( t ) \|_{2}^{2} \nonumber \\
	&\quad{} + \, \eta \sum_{i = 1}^{I} \sum_{ j = 1, j \neq i }^{I} L_{1} \| \bm{v}_{i} (t) - \bm{v}_{j} (t) \|_{2} \| \bm{s}_{i} (t) \|_{2} \nonumber \\
	& \quad{} \text{(From Assumption~\ref{S4-A2})} \nonumber \\
	&\leq \bar{F} \left( \bm{w} ( t ) \right) + \left( - \dfrac{\eta I}{L_{3}} + \dfrac{\eta \delta}{L_{2}} ( I - 1 ) + \dfrac{L_{1}}{2} I \eta^{2} \right) \| \bm{g} ( t ) \|_{2}^{2} \nonumber \\
	&\quad{} + \, \eta I \sum_{i = 1}^{I} L_{1} \| \bm{w} ( t ) - \bm{v}_{i} ( t ) \|_{2} \| \bm{s}_{i} ( t ) \|_{2} \nonumber \\
	&\quad{} + \, \eta \sum_{i = 1}^{I} \sum_{ j = 1, j \neq i }^{I} L_{1} \| \bm{w} (t) - \bm{v}_{j} (t) \|_{2} \| \bm{s}_{i} (t) \|_{2} \label{EQ-SA-B1b} \\
	&\leq \left( - \dfrac{\eta I}{L_{3}} + \dfrac{\eta \delta}{L_{2}} ( I - 1 ) + \dfrac{I L_{1} \eta^{2}}{2} + \dfrac{ ( 3 I - 1 ) \eta L_{1} }{2} \right) \| \bm{g} (t) \|_{2}^{2} \nonumber \\
	& \quad{} + \, \bar{F} \left( \bm{w} ( t ) \right) + \dfrac{ ( 3 I - 1 ) \eta L_{1} }{2} \| \bm{w} (t) - \bm{v}_{i} ( t ) \|_{2}^{2}, \label{EQ-SA-B1c}
	\end{align}
\end{subequations}
where \eqref{EQ-SA-B1b} is derived by
\begin{align}
\lefteqn{\sum_{i = 1}^{I} \sum_{ j = 1, j \neq i }^{I} L_{1} \| \bm{v}_{i} (t) - \bm{v}_{j} (t) \|_{2} \| \bm{s}_{i} (t) \|_{2}} \nonumber \\
&\quad{} + \, \sum_{i = 1}^{I} L_{1} \| \bm{w} ( t ) - \bm{v}_{i} ( t ) \|_{2} \| \bm{s}_{i} ( t ) \|_{2} \nonumber \\
&\leq {\small \sum_{i = 1}^{I} \sum_{ j = 1, j \neq i }^{I} L_{1} \left( \| \bm{w} (t) - \bm{v}_{j} (t) \|_{2} + \| \bm{v}_{i} (t) - \bm{w} (t) \|_{2} \right)  \| \bm{s}_{i} (t) \|_{2}} \nonumber \\
&\quad{} + \sum_{i = 1}^{I} L_{1} \| \bm{w} ( t ) - \bm{v}_{i} ( t ) \|_{2} \| \bm{s}_{i} ( t ) \|_{2} \nonumber \\
&\quad \text{(From the triangular inequality)} \nonumber \\
&\leq \sum_{i = 1}^{I} \sum_{ j = 1, j \neq i }^{I} L_{1} \| \bm{w} (t) - \bm{v}_{j} (t) \|_{2} \| \bm{s}_{i} (t) \|_{2} \nonumber \\
&\quad{} + \, I \sum_{i = 1}^{I} L_{1} \| \bm{v}_{i} (t) - \bm{w} (t) \|_{2} \| \bm{s}_{i} (t) \|_{2}, \nonumber
\end{align}
and \eqref{EQ-SA-B1c} is obtained by
\begin{align} \nonumber
\lefteqn{L_{1} \sum_{i = 1}^{I} \sum_{ j = 1, j \neq i }^{I} \| \bm{w} (t) - \bm{v}_{j} (t) \|_{2} \| \bm{s}_{i} (t) \|_{2}} \nonumber \\
&\quad{} + \, I L_{1} \sum_{i = 1}^{I} \| \bm{w} (t) - \bm{v}_{i} ( t ) \|_{2} \| \bm{s}_{i} ( t ) \|_{2} \nonumber \\
&\leq L_{1} \sum_{i = 1}^{I} \sum_{ j = 1, j \neq i }^{I} \dfrac{\| \bm{w} (t) - \bm{v}_{j} (t) \|_{2}^{2} + \| \bm{s}_{i} (t) \|_{2}^{2}}{2} \nonumber \\
&\quad{} + \, I L_{1} \sum_{i = 1}^{I} \dfrac{\| \bm{w} (t) - \bm{v}_{i} (t) \|_{2}^{2} + \| \bm{s}_{i} (t) \|_{2}^{2}}{2} \nonumber \\
&\quad \text{(By the inequality $\|\bm{a}\|_{2} \|\bm{b}\|_{2} \leq \left( \|\bm{a}\|_{2}^{2} + \|\bm{b}\|_{2}^{2} \right) / 2$)} \nonumber \\
&\leq \dfrac{ ( 3 I - 1 ) L_{1} }{2} \| \bm{g} ( t ) \|_{2}^{2} + \dfrac{ ( 3 I - 1 ) L_{1} }{2} \sum_{i = 1}^{I} \| \bm{w} (t) - \bm{v}_{i} (t) \|_{2}^{2}. \nonumber
\end{align}

Next, to bound $\| \bm{w} (t) - \bm{v}_{i} (t) \|_{2}^{2}$, we have 
\begin{align}
\| \bm{w} ( t ) - \bm{v}_{i} ( t ) \|_{2}^{2} &= \left\| \bm{w} ( t ) - \sum_{ j \in \mathcal{I} } \alpha_{j} \bm{w}_{j} \left( \tau_{i, \, j} ( t ) \right) \right\|_{2}^{2} \nonumber \\
&= \eta^{2} \left\| \sum_{ j \in \mathcal{I} } \alpha_{j} \sum_{\tau = \tau_{i, \, j} ( t )}^{t - 1} \bm{s}_{j} (\tau) \right\|_{2}^{2} \nonumber \\
& \leq \sum_{ j \in \mathcal{I} } \alpha_{j} \eta^{2} \left\| \sum_{\tau = \tau_{i, \, j} ( t )}^{t - 1} \bm{s}_{j} (\tau) \right\|_{2}^{2} \nonumber \\
&\leq \eta^{2} \Gamma \sum_{ j \in \mathcal{I} } \sum_{\tau = t - \Gamma}^{t - 1} \alpha_{j} \left\| \bm{s}_{j} (\tau) \right\|_{2}^{2} \label{EQ-SA-B4a} \\
&\leq \eta^{2} \Gamma \sum_{\tau = t - \Gamma}^{t - 1} \left\| \bm{g} (\tau) \right\|_{2}^{2}, \label{EQ-SA-B4b}
\end{align}
where \eqref{EQ-SA-B4a} and \eqref{EQ-SA-B4b} are due to Jensen's inequality. By virtue of \eqref{EQ-SA-B4b}, we have
\begin{align}
&\bar{F} \left( \bm{w} ( t + 1 ) \right) \nonumber \\
&\leq - \, \left( \dfrac{\eta I}{L_{3}} - \dfrac{\eta \delta}{L_{2}} ( I - 1 ) - \dfrac{I L_{1} \eta^{2}}{2} - \dfrac{ ( 3 I - 1 ) \eta L_{1} }{2} \right) \| \bm{g} ( t ) \|_{2}^{2} \nonumber \\
&\quad{} + \, \bar{F} \left( \bm{w} ( t ) \right) + \dfrac{ ( 3 I - 1 ) \eta^{3} L_{1} \Gamma }{2} \sum_{\tau = t - \Gamma}^{t - 1} \left\| \bm{g} (\tau) \right\|_{2}^{2}. \label{EQ-SA-B5}
\end{align}
Summing up among the $1^{\rm st}, \, \cdots, \, t^{\rm th}$ iterations, we get
\begin{align}
\lefteqn{\bar{F} \left( \bm{w} ( t + 1 ) \right)} \nonumber \\
&\leq \bar{F} \left( \bm{w} ( 0 ) \right) - \sum_{\tau = 0}^{t} \| \bm{g} ( \tau ) \|_{2}^{2} \left( \dfrac{\eta I}{L_{3}} - \dfrac{\eta \delta}{L_{2}} ( I - 1 ) - \dfrac{I L_{1} \eta^{2}}{2} \right. \nonumber \\
&\quad{} -\, \left. \dfrac{ ( 3 I - 1 ) \eta L_{1} }{2} \right) + \dfrac{ ( 3 I - 1 ) \eta^{3} L_{1} \Gamma^{2} }{2} \sum_{\tau = 0}^{t - 1} \left\| \bm{g} (\tau) \right\|_{2}^{2} \nonumber \\
&\leq \bar{F} \left( \bm{w} ( 0 ) \right) - \eta \left( \dfrac{I}{L_{3}} - \dfrac{\delta (I - 1)}{L_{2}} - \dfrac{\left(I (3 + \eta) - 1 \right) L_{1}}{2} \right. \nonumber \\
&\quad{} -\, \left. \dfrac{ (3 I - 1) \eta^{2} L_{1} \Gamma^{2} }{2} \right) \sum_{\tau = 0}^{t} \left\| \bm{g} (\tau) \right\|_{2}^{2}. \nonumber
\end{align}
This completes the proof.

\section{Proof of Corollary~\ref{S5-C1}} \label{SA-C}
From Assumption~\ref{S4-A2}(a), we have $\bar{F} (\bm{w} (t + 1)) \geq 0$, and then \eqref{S5-EQ-8} can be rewritten as
\begin{equation} \label{SA-EQ-C1}
\sum_{\tau = 0}^{t} \left\| \bm{g} (\tau) \right\|_{2}^{2} \leq \dfrac{1}{\eta u (\eta)} \bar{F} \left( \bm{w} ( 0 ) \right) < + \infty, \, t \geq 0
\end{equation}
which implies that 
\begin{equation} \label{EQ-SA-D2}
\lim\limits_{t \rightarrow + \infty} \bm{g} (t) = 0, \, \lim\limits_{t \rightarrow + \infty} \bm{s}_{i} (t) = 0, \, \forall i \in \mathcal{I}.
\end{equation}
By using \eqref{S4-EQ6}, we deduce that
\begin{equation} \label{EQ-SA-D5}
\bm{w}_{i} (t) = \bm{w}_{i} ( \tau_{j, \, i} ( t ) ) + \eta \sum_{\tau = \tau_{j, \, i} ( t )}^{t - 1} \bm{s}_{i} (\tau), \, \forall i \in \mathcal{I}.
\end{equation}
As $t - \tau_{j, \, i} (t) \leq \Gamma$, we obtain $\lim\limits_{t \rightarrow + \infty} \tau_{j, \, i} (t) = t \rightarrow + \infty$, and 
\begin{align} 
&\lim\limits_{t \rightarrow + \infty} \left\| \sum_{\tau = \tau_{j, \, i} ( t )}^{t - 1} \bm{s}_{i} (\tau) \right\|_{2} \leq \lim\limits_{t \rightarrow + \infty} \sum_{\tau = \tau_{j, \, i} ( t )}^{t - 1} \left\| \bm{s}_{i} (\tau) \right\|_{2} \nonumber \\
&\leq \Gamma \lim\limits_{t \rightarrow + \infty} \left\| \bm{s}_{i} (t) \right\|_{2} = 0, \, \forall i \in \mathcal{I}, \label{EQ-SA-D6}
\end{align}
where \eqref{EQ-SA-D6} is obtained by \eqref{EQ-SA-D2}. Using \eqref{EQ-SA-D6}, we have
\begin{align}
&\lim\limits_{t \rightarrow + \infty} \| \bm{w}_{i} (t) - \bm{w}_{i} ( \tau_{j, \, i} ( t ) ) \|_{2} = 0, \nonumber \\
&\lim\limits_{t \rightarrow + \infty} \bm{w}_{i} (t) = \lim\limits_{t \rightarrow + \infty} \bm{w}_{i} ( \tau_{j, \, i} ( t ) ), \, \forall i \in \mathcal{I}.
\end{align}	
Moreover, we have
\begin{align} 
&\| \bm{w} ( t ) - \bm{v}_{i} ( t ) \|_{2} = \left\| \bm{w} ( t ) - \sum_{ j \in \mathcal{I} } \alpha_{j} \bm{w}_{j} \left( \tau_{i, \, j} ( t ) \right) \right\|_{2} \nonumber \\
&= \eta \left\| \sum_{ j \in \mathcal{I} } \alpha_{j} \sum_{\tau = \tau_{i, \, j} ( t )}^{t - 1} \bm{s}_{j} (\tau) \right\|_{2}, \, \forall i \in \mathcal{I}. \label{EQ-SA-D7}
\end{align}
Combining \eqref{EQ-SA-D6} and \eqref{EQ-SA-D7} concludes that
\begin{equation} \nonumber
\lim\limits_{t \rightarrow + \infty} \| \bm{w} (t) - \bm{v}_{i} (t) \|_{2} = 0, \, \lim\limits_{t \rightarrow + \infty} \bm{w} (t) = \lim\limits_{t \rightarrow + \infty} \bm{v}_{i} (t).
\end{equation}

\bibliographystyle{IEEEtran}
\bibliography{ref}

\begin{IEEEbiography}
	[{\includegraphics[width=1in, height=1.25in, clip, keepaspectratio]{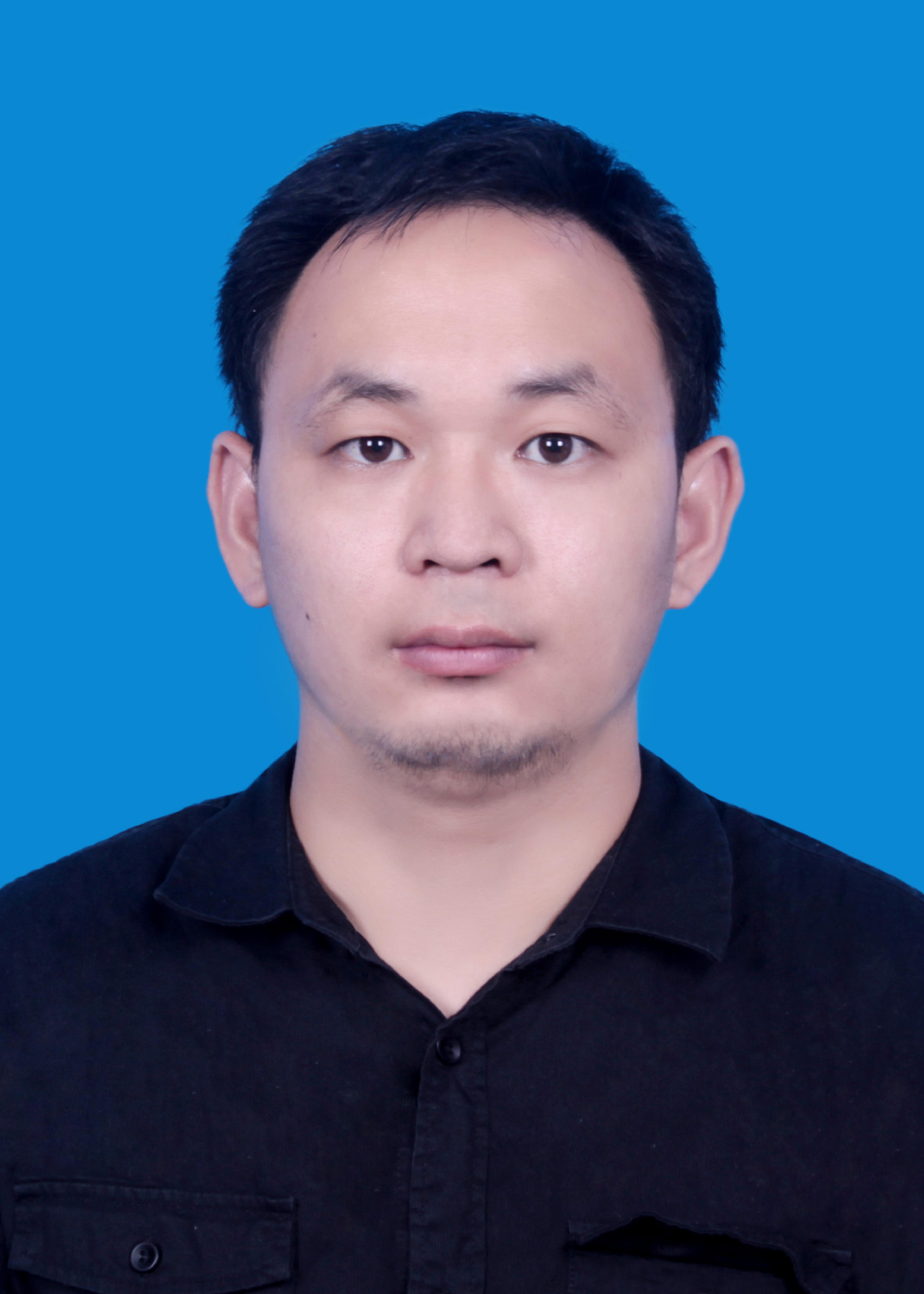}}]{Haihui Xie} received the B.S. degree and the M.S. degree in photonic and electronic engineering from Fujian Normal University, Fuzhou, China, in 2014 and 2016, respectively, and the Ph.D. degree in information and communication engineering from Sun Yat-sen University, Guangzhou, China, in 2023. He is currently a Lecturer at the School of Fujian Agriculture and Forestry University (FAFU), Fuzhou. His research interests include edge learning, optimization, and wireless communications.
\end{IEEEbiography}

\vfill

\begin{IEEEbiography}
	[{\includegraphics[width=1in, height=1.25in, clip, keepaspectratio]{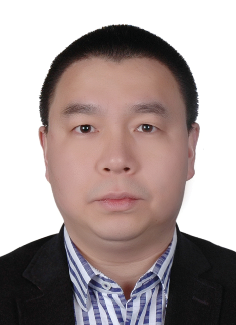}}]{Minghua Xia} (Senior Member, IEEE) received the Ph.D. degree in Telecommunications and Information Systems from Sun Yat-sen University, Guangzhou, China, in 2007.
	
	From 2007 to 2009, he was with the Electronics and Telecommunications Research Institute (ETRI) of South Korea, Beijing R\&D Center, Beijing, China, where he worked as a member and then as a senior member of the engineering staff. From 2010 to 2014, he was in sequence with The University of Hong Kong, Hong Kong, China; King Abdullah University of Science and Technology, Jeddah, Saudi Arabia; and the Institut National de la Recherche Scientifique (INRS), University of Quebec, Montreal, Canada, as a Postdoctoral Fellow. Since 2015, he has been a Professor at Sun Yat-sen University. Since 2019, he has also been an Adjunct Professor with the Southern Marine Science and Engineering Guangdong Laboratory (Zhuhai). His research interests are in the general areas of wireless communications and signal processing.
\end{IEEEbiography}

\vfill

\begin{IEEEbiography}
	[{\includegraphics[width=1in, height=1.25in, clip, keepaspectratio]{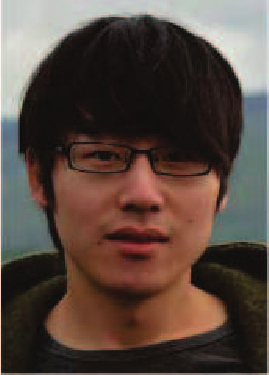}}]{Peiran Wu} (Member, IEEE) received a Ph.D. degree in electrical and computer engineering from The University of British Columbia (UBC), Vancouver, Canada, in 2015.
	
	From October 2015 to December 2016, he was a Post-Doctoral Fellow at UBC. In the Summer of 2014, he was a Visiting Scholar with the Institute for Digital Communications, Friedrich-Alexander-University Erlangen-Nuremberg (FAU), Erlangen, Germany. Since February 2017, he has been with Sun Yat-sen University, Guangzhou, China, where he is currently an Associate Professor. Since 2019, he has been an Adjunct Associate Professor with the Southern Marine Science and Engineering Guangdong Laboratory, Zhuhai, China. His research interests include mobile edge computing, wireless power transfer, and energy-efficient wireless communications. 
	
	Dr. Wu was a recipient of the Fourth-Year Fellowship in 2010, the C. L. Wang Memorial Fellowship in 2011, the Graduate Support Initiative (GSI) Award from UBC in 2014, the German Academic Exchange Service (DAAD) Scholarship in 2014, and the Chinese Government Award for Outstanding Self-Financed Students Abroad in 2014.
\end{IEEEbiography}

\vfill

\begin{IEEEbiography}
	[{\includegraphics[width=1in, height=1.25in, clip, keepaspectratio]{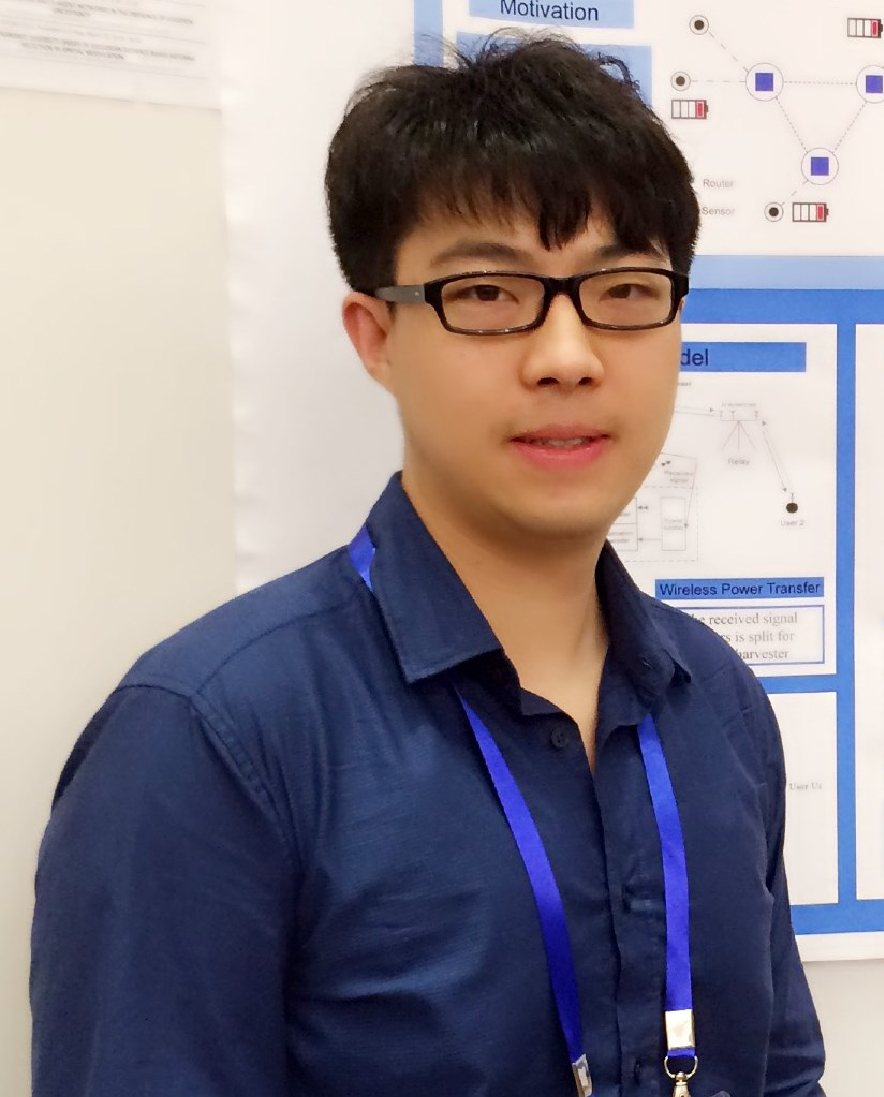}}]{Shuai Wang} (Member, IEEE)
	received the B.Eng. and M.Eng. degrees from the Beijing University of Posts and Telecommunications (BUPT) in 2011 and 2014, respectively, and the Ph.D. degree from the University of Hong Kong (HKU) in 2018. He is now an Associate Professor with the Shenzhen Institute of Advanced Technology (SIAT), Chinese Academy of Sciences. His research interests lie at the intersection of autonomous systems and wireless communications. He has published more than 80 papers in top journals and conferences. He received the Best Paper Awards from IEEE ICC in 2020, IEEE ComSoc SPCC Committee in 2021, and IEEE ICDCSW in 2023. He is currently an Associate Editor for IEEE Open Journal of the Communications Society.
\end{IEEEbiography}

\vfill

\begin{IEEEbiography}
	[{\includegraphics[width=1in, height=1.25in, clip, keepaspectratio]{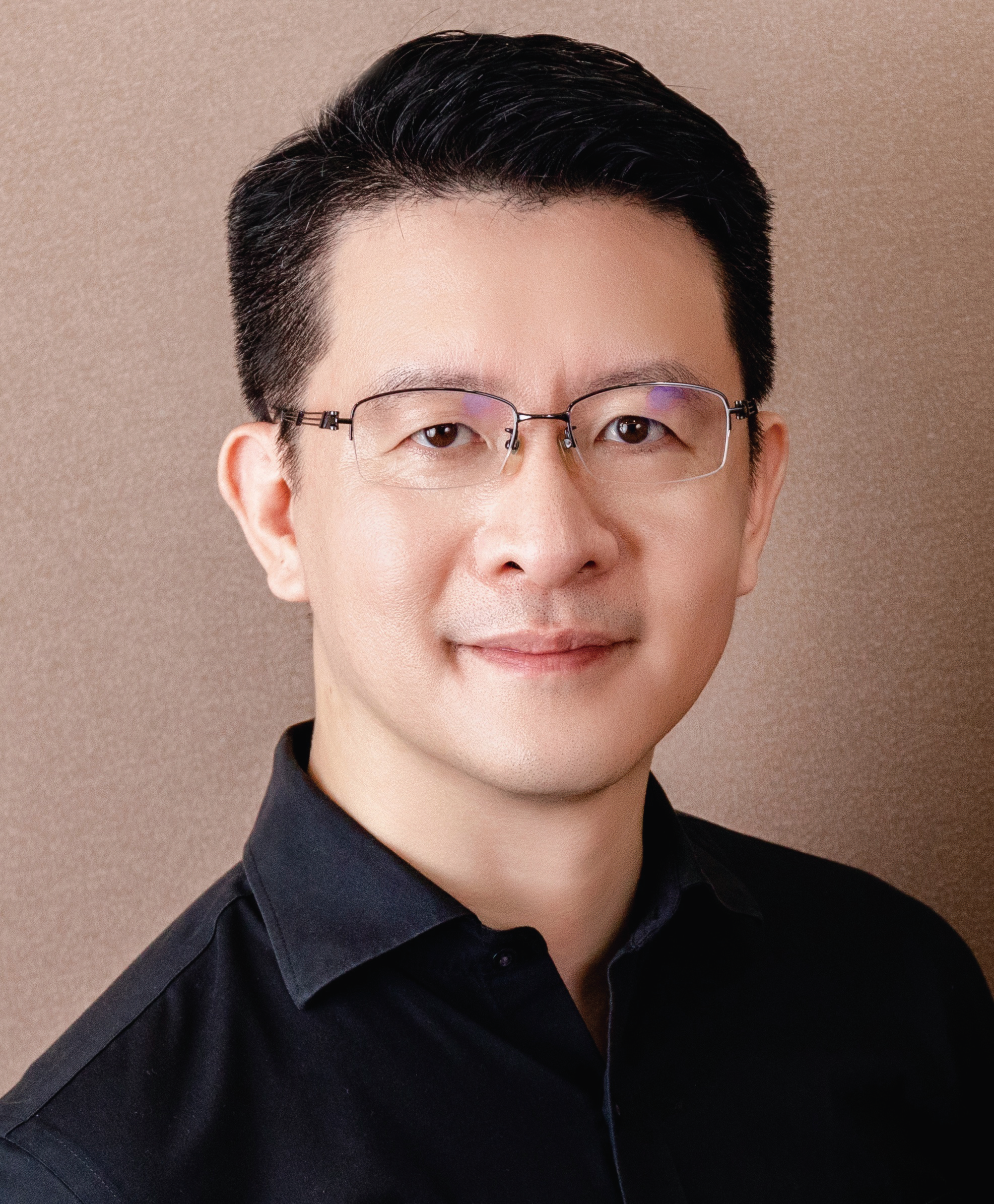}}]{Kaibin Huang} (Fellow, IEEE) received the B.Eng. and M.Eng. degrees from the National University of Singapore and the Ph.D. degree from The University of Texas at Austin, all in electrical engineering. He is a Professor and the Deputy Head at the Dept. of Electrical and Electronic Engineering, The University of Hong Kong, Hong Kong. He received the IEEE Communication Society’s 2021 Best Survey Paper, 2019 Best Tutorial Paper, 2019 and 2023 Asia–Pacific Outstanding Paper, 2015 Asia–Pacific Best Paper Award, and the best paper awards at IEEE GLOBECOM 2006 and IEEE/CIC ICCC 2018. He received the Outstanding Teaching Award from Yonsei University, South Korea, in 2011. He has been named as a Web-of-Science Highly Cited Researcher in 2019-2023. He is a member of the Engineering Panel of Hong Kong Research Grants Council (RGC) and a RGC Research Fellow. He served as the Lead Chair for the Wireless Communications Symposium of IEEE Globecom 2017 and the Communication Theory Symposium of IEEE GLOBECOM 2023 and 2014, and the TPC Co-chair for IEEE PIMRC 2017 and IEEE CTW 2023 and 2013. He is also an Executive Editor of IEEE Transactions on Wireless Communications, and an Area Editor for both IEEE Transactions on Machine Learning in Communications and Networking and IEEE Transactions on Green Communications and Networking. Previously, he served on the Editorial Boards for IEEE Journal on Selected Areas in Communications and IEEE Wireless Communication Letters. He has guest edited special issues of IEEE Journal on Selected Areas in Communications, IEEE Journal of Selected Areas in Signal Processing, and IEEE Communications Magazine. He is also a Distinguished Lecturer of the IEEE Communications Society and the IEEE Vehicular Technology Society.
\end{IEEEbiography}

\vfill
\end{document}